\documentclass{emulateapj}

\newcommand{\lya} {Ly$\alpha$}

\newcommand{\K}   {\mathrm{K}}
\newcommand{\SB}  {\mathrm{ergs\ s^{-1}\ cm^{-2}\ arcsec^{-2}}}
\newcommand{\eV}  {\mathrm{eV}}

\newcommand{\sss} {\;\!\scriptscriptstyle}

\newcommand{\msun}  {M_{\sun}}
\newcommand{\hlya}  {\ion{H}{1}  $\lambda 1216$}
\newcommand{\helya} {\ion{He}{2} $\lambda  304$}
\newcommand{\heha}  {\ion{He}{2} $\lambda 1640$}

\newcommand{\hi}    {\ion{H}{1} }
\newcommand{\hii}   {\ion{H}{2} }
\newcommand{\hei}   {\ion{He}{1} }
\newcommand{\heii}  {\ion{He}{2} }
\newcommand{\heiii} {\ion{He}{3} }

\newcommand{\Hlya}  {H$\;${\footnotesize\rmfamily I} $\lambda 1216$}

\newcommand{\Heha}  {He$\;${\footnotesize\rmfamily II} $\lambda 1640$}
\newcommand{\Hi}    {H$\;${\footnotesize\rmfamily I} }

\newcommand{\Heii}  {He$\;${\footnotesize\rmfamily II} }

\slugcomment{Accepted to \apj}
\shorttitle{\heii Cooling Lines}
\shortauthors{Yang et al.}

\begin{document}

\title{Probing Galaxy Formation with \heii Cooling Lines}
\author{Yujin Yang, Ann I. Zabludoff, Romeel Dav{\' e}, Daniel J. Eisenstein, Philip A. Pinto}
\affil{Steward Observatory, University of Arizona, Tucson, AZ 85721}
 
\author{Neal Katz}
\affil{Astronomy Department, University of Massachusetts, Amherst, MA 01003}

\author{David H. Weinberg}
\affil{Department of Astronomy, The Ohio State University, Columbus, OH 43210}

\author{Elizabeth J. Barton}
\affil{Department of Physics and Astronomy, University of California, Irvine, Irvine, CA 92697}

\begin{abstract}
Using high resolution cosmological simulations, 
we study hydrogen and helium gravitational cooling radiation.
We focus on the \heii cooling lines, 
which arise from gas with a different temperature history 
($T_{\rm{max}} \sim 10^5\ \K$) than \hi line emitting gas.
We examine whether three major atomic cooling lines, 
\hlya, \heha\ and \helya, are observable, finding that 
\lya\ and \heha\ cooling emission at $z=2-3$ are potentially detectable 
with deep narrow band ($R>100$) imaging and/or spectroscopy from the ground.
While the expected strength of \hlya\ cooling emission depends strongly 
on the treatment of the self-shielded phase of the IGM in the simulations,
our predictions for the \heha\ line are more robust
because the \heii emissivity is negligible below $T\sim10^{4.5\ } \K$ 
and less sensitive to the UV background.
Although \heha\ cooling emission is fainter than \lya\ 
by at least a factor of 10 and, unlike \lya, 
might not be resolved spatially with current observational facilities,
it is more suitable to study gas accretion in the galaxy formation process
because it is optically thin  
and less contaminated by the recombination lines from star-forming galaxies.
The \heha\ line can be used to distinguish among mechanisms
for powering the so-called ``\lya\ blobs'' 
--- including gravitational cooling radiation, 
photoionization by stellar populations, and starburst-driven superwinds --- 
because 
(1) \heha\  emission is limited to very low metallicity
($\mathrm{log}(Z/Z_{\sun}) \la -5.3$) and Population III stars, and
(2) the blob's kinematics are probed unambiguously through the \heii line width,
which, for cooling radiation, is narrower ($\sigma < 400\ \mathrm{km\ s^{-1}}$)
than typical wind speeds.
\end{abstract}

\keywords{
cosmology: theory ---
galaxies: formation ---
intergalactic medium
}

\section{INTRODUCTION}
\label{sec:intro}

Galaxies grow partly by accretion of gas from the surrounding
intergalactic medium and partly by mergers with other galaxies.
Observational studies of galaxy assembly have focused primarily
on merger rates, which can be measured indirectly by counting
close pairs and merger remnants.  However, all the mass that enters
the galaxy population ultimately does so by accretion --- mergers
can only redistribute this mass from smaller systems to larger
systems.  Furthermore, numerical simulations predict that even
large galaxies grow primarily by smooth gas accretion rather
than by cannibalism of smaller objects \citep{Murali,Keres}.
Gas shock-heated to the virial temperature of a typical dark
matter halo would radiate most of its acquired gravitational
energy in the soft X-ray continuum, making individual sources 
very difficult to detect, especially at high redshift.
However, Fardal et al. (\citeyear{Fardal}, hereafter F01)
show that much of the gas that enters galaxies in hydrodynamic
cosmological simulations never heats to high temperatures
at all, and that it therefore channels a substantial fraction
of its cooling radiation into atomic emission lines, 
especially \hi \lya.  F01 and \cite{Haiman} suggested 
that extended ``\lya\ blobs''
\citep[e.g.,][]{Keel,Steidel,Francis,Matsuda,Dey},
with typical sizes of $10-20\arcsec$ and line luminosities
$L_{\rm{Ly\alpha}}\sim10^{44}\ \rm{ergs\ s^{-1}}$, might be signatures
of cooling radiation from forming galaxies.
\cite{Furlanetto:05} have also investigated predictions for \lya\ 
cooling radiation from forming galaxies in hydrodynamic simulations.

In this paper, we investigate other aspects of cooling radiation
from forming galaxies, in particular the potentially detectable
radiation in the \helya\ (``\lya'') and \heha\ (``H$\alpha$'') lines
of singly ionized helium.  While challenging, the successful detection
\heii line emission would complement \hi \lya\ measurements
in at least three ways.  First, because \hi and \heii line cooling rates
peak at different temperatures ($T\sim 10^{4.3}\,\K$ vs. $T\sim 10^5\,\K$),
measurements of both lines could constrain the physical conditions
of the emitting gas.  Recent theoretical studies imply that
``cold mode'' accretion, in which the maximum gas temperature is
well below the halo virial temperature, is a ubiquitous and 
fundamental feature of galaxy formation 
(F01; \citealt{Katz:03,Birnboim,Keres,Dekel}), a view anticipated
by the early analytical work of \cite{Binney}.
Simultaneous measurements of \heii and \hi emission from different
types of galaxies could eventually test detailed predictions for
the temperature distribution of cooling gas.

The second advantage of the \heha\ line is that it should be optically
thin, allowing a straightforward interpretation of the observed
emission in terms of the spatial distribution and kinematics of the
cooling gas.  In contrast, the radiative transfer effects on the \hi
\lya\ emission from accreting gas are more complicated, an issue that
we will investigate in future work (J.\ Kollmeier et al., in
preparation; see also \citealt{Cantalupo}).

Third, the different temperature dependence and absence of radiative
transfer effects in \heii lines could help distinguish cooling
radiation from alternative explanations of ``\lya\ blobs,'' such as
emission from collisionally ionized gas in galactic superwinds
\citep{Taniguchi} or from gas photoionized by young stellar
populations.  For example, only the lowest metallicity stars
($\mathrm{log}\,Z/Z_{\sun} \la -5.3$) have hard enough spectra to
ionize \heii to \heiii, so stellar photoionization will generally not
produce \heii line emission.

It is possible that \heii cooling emission from forming galaxies 
at high redshift has already been detected.
One such example is the broad \heha\ line
in the composite spectrum of Lyman break galaxies (LBGs) from \citet{Shapley}.
The composite \heii line shows a rather broad line width 
(FWHM$\sim$ 1500 km s$^{-1}$), 
which is a possible signature of Wolf-Rayet stars.
However, it is difficult to reproduce the strength of the \heii line
via stellar population models with reasonable parameters \citep{Shapley}.
We show in this paper that the \heii cooling emission around individual 
galaxies is detectable, which suggests it might be fruitful 
to search for \heha\ cooling emission in the outer parts of individual LBG's.

The next section describes
our simulations and the radiative cooling mechanisms 
that lead to \hlya, \heha, and \helya\ line emission.
We present \hlya\ and \heha\ cooling maps in \S\ref{sec:map}
and discuss the properties of the cooling radiation
sources in \S \ref{sec:properties}.
In \S \ref{sec:detectability}, 
we examine the detectability of the three major cooling lines
in the far ultraviolet and optical and 
discuss the best observational strategies.
In \S \ref{sec:dis}, we discuss 
mechanisms other than gravitational cooling radiation
that can generate extended \lya\ emission and describe 
how the \heii cooling line might help us discriminate among those mechanisms.
We summarize our conclusions in \S \ref{sec:conclusion}.

\section{SIMULATIONS AND COOLING RADIATION}
\label{sec:sim}
We use Parallel TreeSPH simulations \citep{Dave:97}
including the effects of radiative cooling, star formation, thermal feedback, 
and a spatially uniform metagalactic photoionizing background.  
We analyze two simulations: 
one with a cubic volume of $11.111\ h^{-1}$ Mpc (comoving) on a side and
a spatial resolution of $1.75\ h^{-1}$ kpc (comoving; equivalent Plummer softening),
the other with a cubic volume of $22.222\ h^{-1}$ Mpc on a side
and $3.5\ h^{-1}$ kpc resolution.
Hereafter, we refer to these two simulations 
as the 11 Mpc and 22 Mpc simulations, respectively.
The simulations consist of $128^3$ dark matter particles and $128^3$ gas particles, 
giving a mass resolution of 
$m_{\mathrm{SPH}}=1.3\times10^7 \msun$ and $m_{\mathrm{dark}}=10^8 \msun$ 
for the 11 Mpc simulation, and 
$m_{\mathrm{SPH}}=1.1\times10^8 \msun$ and $m_{\mathrm{dark}}=7.9\times10^8 \msun$ 
for the 22 Mpc simulation.
We adopt a $\Lambda$CDM cosmology with the parameters 
$\Omega_\mathrm{M}=0.4$, $\Omega_{\Lambda}=0.6$, $\Omega_{b} h^2=0.02$, 
and $H_0=65\ \mathrm{km\ s^{-1}\ Mpc^{-1}}$.
In our calculation of cooling emission lines, we basically follow 
the radiative cooling processes described in \citet[][hereafter KWH]{Katz}. 
Here, we briefly summarize the cooling processes that can contribute to 
line emission. 

The four underlying assumptions of radiative cooling are:
primordial composition, ionization equilibrium, an optically thin gas, 
and a spatially uniform radiation field. 
In the simulations, we adopt $X = 0.76$ and $Y = 0.24$, 
where $X$ and $Y$ are the hydrogen and helium abundance by mass.
The abundances of each ionic species 
($\mathrm{H^0}$, $\mathrm{H^+}$, $\mathrm{He^0}$, $\mathrm{He^+}$, $\mathrm{He^{++}}$)
are solely determined by assuming that the primordial plasma is optically thin 
and in ionization equilibrium (but not in thermal equilibrium).
The functional forms of the temperature-dependent recombination rates, 
collisional ionization rates, collisional excitation rates, 
and the rate equations are given in \S {3} of KWH and the tables therein.
The uniform photoionizing UV background is taken from \citet{Haardt}.

In the next section, we describe the \lya\ and \heii cooling curves 
under the optically thin gas assumption, 
the assumption used in our simulations.
However, in the high density regime, 
a gas cloud becomes dense 
enough to shield the central part of itself from the UV background, 
i.e., becomes self-shielded.
Therefore, the actual emissivity of this self-shielded 
(or condensed phase) gas is highly uncertain.
We discuss below how to treat this phase of the IGM 
to derive better estimates of the cooling radiation.
Once we generate the \lya\ and \heii cooling emissivities,
the cooling radiation is determined 
by how many gas particles populate a certain range of temperature and density,
where we appeal to the high-resolution cosmological simulations mentioned above. 

\begin{figure*}[t]
\epsscale{1.15}
\plotone{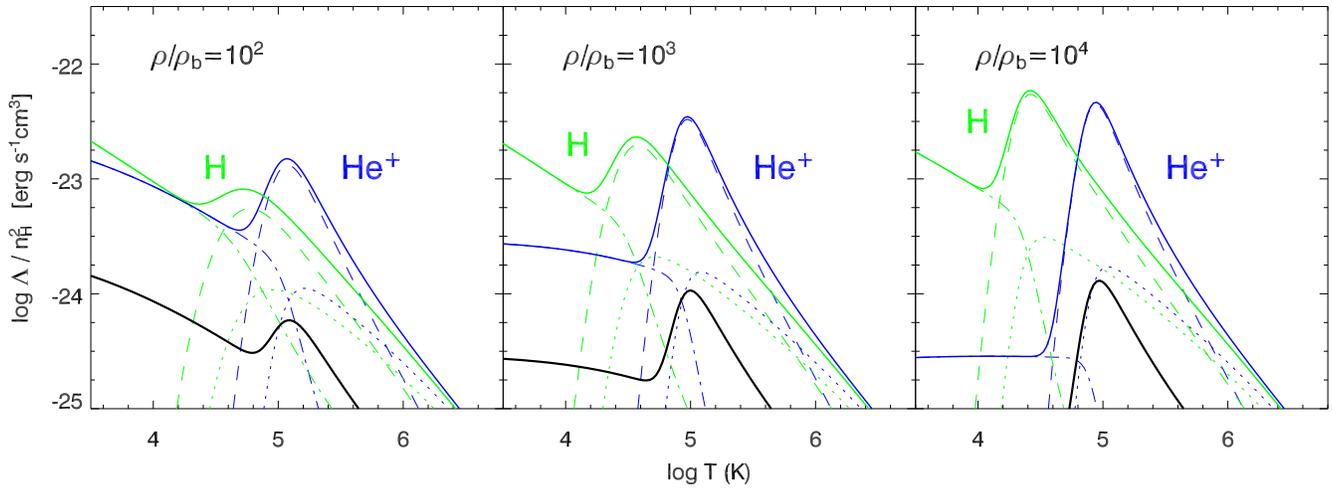}
\caption{
Normalized line emissivity $\mathrm{log \Lambda}/n^2_{\mathrm{H}}$
as a function of temperature for a primordial plasma at
densities $\rho/\bar{\rho_b} = 10^2$, $10^3$, and $10^4$
(\emph{from left to right}) in the presence of a UV ionizing background at $z=3$.
In each panel, the dashed lines represent the collisional excitation cooling rates.
The dot-dashed lines and dotted lines denote the recombination rates
owing to photoionization and the collisional ionization, respectively.
The solid lines represent the total line cooling rates of hydrogen and helium.
The bold solid lines below the \Heii cooling curves
represent the \Heha\ line emissivity.
Compared with H I,
the cooling rates of \Heii owing to the UV ionizing background
become significantly weaker as the gas density increases.
}
\label{fig:cooling}
\end{figure*}

\subsection{Cooling Curves}
\label{sec:CoolingCurves}
Among the various radiative cooling processes, 
only two can produce 
\hi \lya\ (1216\AA), 
\heii \lya\ (340\AA), and 
\heii Balmer $\alpha$ (1640\AA) photons:
the recombination cascades of a free electron and 
the collisional excitation of a bound electron to an excited state 
followed by radiative decay.
The dominant cooling mechanism is the collisional excitation of 
neutral hydrogen and singly ionized helium, 
which have their peaks at temperatures of
$T \sim 10^{4.3\ }\K$ and $\sim 10^{5\ }\K$, respectively.
Figure \ref{fig:cooling} shows the \hlya\ and \helya\ emissivities 
for gas of different densities in ionization equilibrium 
in the presence of a metagalactic photoionizing background.
The dashed lines represent the collisional excitation cooling lines
of neutral hydrogen and singly ionized helium.
The dot-dashed lines and dotted lines denote the recombination lines
for these two species due to photoionization and collisional ionization, 
respectively.
The solid lines represent the total \lya\ cooling rates of hydrogen and helium.
Below $T \sim 10^{4\ }\K$,
collisions with free electrons are not energetic enough to raise bound electrons
to upper levels or to ionize the neutral hydrogen, so the collisional
cooling rate of hydrogen drops quickly below this temperature. 
For singly ionized helium, the collisional cooling rate
drops to virtually zero below $T \sim 10^{4.6\ }\K$.
Therefore, below $T \sim 10^{4\ }\K$, photoionization 
by the background UV spectrum (heating) and the following recombination
(cooling) 
is the dominant source for cooling line emission of the primordial plasma.

In the presence of photoionization, the cooling curve depends
on the density as shown in Figure \ref{fig:cooling}, 
because the equilibrium abundances of each species depend on the density. 
While the collisional ionization rate per unit volume scales as $n^2_{\mathrm{H}}$, 
the photoionization rate per volume is proportional to only $n_{\mathrm{H}}$.
In low density gas ($\rho/\bar{\rho_b} \lesssim 10^2$), 
a significant fraction of \hi and \heii is photoionized and 
their abundances are mainly determined by the photoionization process, 
so the collisional excitation feature is relatively weak. 
As the gas density increases ($\rho/\bar{\rho_b} \approx 10^{3}-10^{4}$), 
however, 
collisional excitation becomes more important and 
the cooling curves approach pure collisional equilibrium, 
the so-called coronal equilibrium.
The notable difference between the cooling curves of hydrogen and 
singly ionized helium is that the cooling rates of \heii owing to
photoionization become significantly weaker as the gas density increases.
If we assume that the medium is optically thin to the ionizing background
--- i.e., not self-shielded ---
hydrogen is almost fully ionized over the entire
temperature range even at high densities, 
but $\sim$90\% of the helium is in a singly ionized state 
below $T \sim 10^{4.8\ }\K$ at high densities.
Thus the assumption that the gas is optically thin everywhere is roughly valid 
for the calculation of the \heii line fluxes, while it is poor for hydrogen.
The correction for the self-shielding of hydrogen
will be discussed in the next section.

We estimate the \heha\ flux from the \helya\ flux by considering
the ratio of \heha\ to \helya\ in the recombination cascades and 
in the collisional excitations, respectively. 
The thick solid lines below the \heii cooling curves in Figure \ref{fig:cooling} 
represent our estimate of the \heha\ line emissivity.
Below $T \sim 10^{5\ }\K$, the optical depth of \helya\ is so large that 
case-B recombination is a good approximation. 
Even though the \lya\ optical depth is extremely large, 
the population of the $2p$ and $2s$ states is always much smaller 
than that of the $1s$ state,
because the de-excitation time for level transitions is very short
($A_{2p1s} \approx 10^{10} s^{-1}$).
One might be concerned whether the population of the $2s$ state 
is large because of the forbidden transition ($2s\rightarrow1s$),
but the two photon decay process is fast enough 
to de-populate $2s$ electrons ($A_{2s1s} \simeq 8.22 Z^6 s^{-1}$).
Therefore, Balmer lines are always optically thin.
We adopt $F^{\rm{rec}}_{1640}/{F^{\rm{rec}}_{304}} \simeq 10\%$ by extrapolating 
the case-B values of \citet{Storey} to the low density limit.
For collisional excitation, we estimate the \heha\ flux using 
\begin{equation}
	\frac{F^{\rm{coll}}_{1640}}{F^{\rm{coll}}_{304}} \simeq 
	\frac{C_{1s3s} + C_{1s3p} + C_{1s3d}}{C_{1s2p} + C_{1s3s} + C_{1s3d}}
	\frac{h\nu_{1640}}{h\nu_{304}} ,
\end{equation}
where $C_{i j}$ is the collisional excitation rate from the $i$ to the $j$ state.
We adopt the $C_{i j}$'s from \citet{Aggarwal}.
$F^{\rm{coll}}_{1640}/F^{\rm{coll}}_{304}$ is roughly $2-4\%$ 
in the temperature range of $10^5<T<10^{5.7}$ 
where \helya\ collisional excitation cooling is dominant.
In summary, the \heha\ flux is calculated by 
\begin{eqnarray}
   && \varepsilon_{{\mathrm{He\sss II\ \lambda 1640}}} = \nonumber \\
   && f_{\mathrm{rec}} n_e n_{\mathrm{He^{++}}}
   \alpha^{\mathrm{rec}}_{\mathrm{He^+}} h \nu_{\mathrm{Ly\alpha}} +
   f_{\mathrm{coll}} n_e n_{\mathrm{He^+}} C_{12} h \nu_{\mathrm{Ly\alpha}} ,
\end{eqnarray}
where $\alpha^{\mathrm{rec}}_{\mathrm{He^+}}$ is the recombination rate
into $n\geqslant2$ states of \ion{He}{2}, 
and by assuming $f_{\mathrm{rec}} \approx 10\%$ and $f_{\mathrm{coll}} \approx 2-4\%$.

\subsection{Self-shielding Correction}
\label{sec:self}
A major difference between our work and that of \citet{Fardal}
is that our simulations include a uniform UV background radiation field 
\citep[see also][]{Furlanetto:05}.
However, because even state-of-art cosmological simulations like ours
do not include radiative transfer,
the self-shielded phase of the gas at high column densities is not treated properly.
When the gas is heated to high temperature ($T \sim 10^5-10^6\ \K$) 
by falling into the forming galaxy's halo, 
the gas is mostly ionized, so it is reasonable
to assume that the gas is optically thin to the uniform UV background.
Subsequently, 
when the gas cloud starts losing its thermal energy via cooling radiation, 
its neutral column density becomes sufficiently high that 
the metagalactic UV radiation cannot penetrate the surrounding gas, 
and the cloud becomes self-shielded.

Once the supply of ionizing photons is shut off, 
what happens to the self-shielded high column density clouds?
First, the ionization states will achieve collisional ionization equilibrium, 
where the emissivity is determined solely by 
collisional ionization      
and collisional excitation. 
Second, because the cooling emissivity is boosted by these processes,
the self-shielded cloud will cool more rapidly to $T\sim10^4\ \K$ than 
in the presence of heating by ionizing photons.
Below this temperature, 
the subsequent cooling is dominated by metal lines (if there are metals).
Stars ultimately form from this cold gas.

Because our simulations do not include the time evolution of 
the self-shielded gas or metal-line cooling,
it is not clear how long the gas particles stay in the self-shielded phase
and emit in collisional ionization equilibrium.
The \lya\ emissivities shown in Figure \ref{fig:cooling} become unreliable 
in this self-shielded regime.
Thus we apply a \emph{pseudo} self-shielding correction 
to the high density gas particles to correct their emissivities. 
This correction is not rigorous; 
to properly calculate the emissivity of the self-shielded phase of the IGM,
one should incorporate a radiative transfer calculation 
that includes non-uniform and anisotropic UV radiation fields.
A different prescription for the self-shielded phase is definitely possible.
For example, \citet{Furlanetto:05} consider two extreme cases:
1) adopting zero emissivity and 
2) using the collisional ionization equilibrium emissivity 
for the self-shielded phase.
Our self-shielding correction scenario described below lies
between these two extremes. 

To apply the self-shielding correction, we first define 
the ``local'' optical depth for each gas particle,
$\tau_{\rm local}(\nu) = \sum_i n_i \sigma_i (\nu)\,(\alpha l)$, 
where $n_i$ and $\sigma_i$ are the number densities and 
the photoionization cross sections of each species 
(H$^0$, He$^0$, He$^+$), respectively. 
The ``local'' size of the gas cloud $l$ 
--- the length that corresponds to the volume 
that the gas particle would occupy in space ---
is defined as $(m_{\rm gas}/\rho)^{1/3}$.
\footnote{
Clumping inside a gas particle and/or among gas particles could be 
approximated using a free parameter $\alpha$ such that 
$\alpha l$ represents the effective geometrical edge-to-center distance 
of the gas cloud. 
For example, $\alpha_{\rm sphere} = (3/4\pi)^{1/3}$ is given for a single
spherical gas cloud, whereas $\alpha=2$ corresponds to the clumping of 
$(\alpha/\alpha_{\rm sphere})^{3} \simeq 34$ gas particles.
The value of $\alpha$ should vary from one particle to another,
but we adopt $\alpha = 1$ throughout the paper as a fiducial value.
The over-density where the self-shielding 
occurs also depends on the choice of the edge-to-center distance $\alpha l$
(and on the redshift). 
However, we find that the effects of adopting $\alpha=0.5-2.0$
is insignificant.
}
For each gas particle, the UV background spectrum $J(\nu)$ is attenuated 
using this local optical depth, i.e. $J(\nu) e^{-\tau(\nu)}$, and
new photoionization/heating rates and equilibrium number densities are calculated.
We then determine a new $\tau_{\rm local}(\nu)$ from these values and 
iterate this procedure until the photoionization rates and 
the optical depths converge.
We use these final ionization/heating rates 
to calculate the \lya\ and \heii emissivity of each gas particle.
This modified emissivity for each gas particle
is what we will refer to as the \emph{self-shielding correction case}.

\begin{figure}
\epsscale{1.1}
\plotone{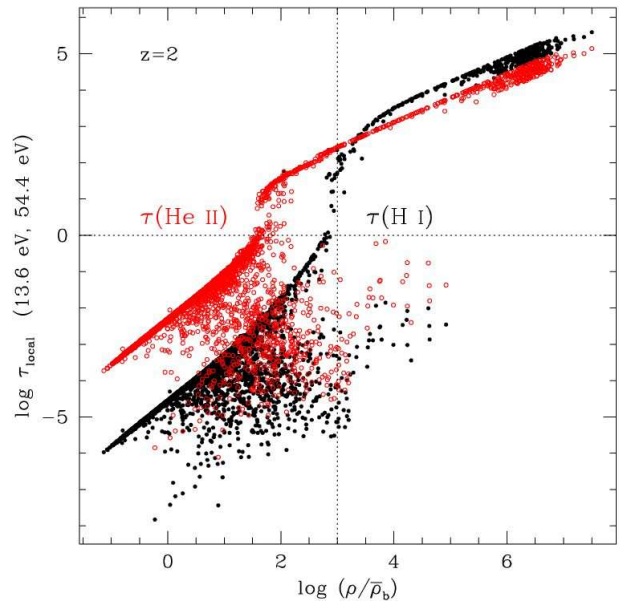}
\caption{
The local optical depth of each gas particle
as a function of over-density at $z=2$.
Solid and open circles indicate the optical depth
at the H I (13.6 eV) and He II (54.4 eV) ionization edges, respectively.
The horizontal and vertical lines represent
$\tau = 1$ and $\rho/\bar{\rho}_b \simeq 10^3$, respectively.
Note that \Hi optical depth increases abruptly at $\rho/\bar{\rho}_b \simeq 10^3$
and the gas becomes self-shielded.
}
\label{fig:tau}
\end{figure}

In Figure \ref{fig:tau}, we show the local optical depths 
of each gas particle in the final equilibrium state as a function of over-density. 
For the 22 Mpc simulation at $z=2$, 
we show $\tau$(\ion{H}{1}) and $\tau$(\ion{He}{2}), 
the optical depth at the \hi(13.6 eV) and \heii(54.4 eV) ionization edges, 
respectively.
As indicated by the dotted lines, 
the \hi optical depth increases abruptly 
from $\tau \simeq 1$ to $\tau=10-100$ at $\rho/\bar{\rho}_b = 10^3$.
Therefore, the optically thin UV background assumption is valid 
below $\rho/\bar{\rho}_b = 10^3$, 
but the gas becomes self-shielded quickly 
above this over-density.
%
Because the transition from the optically thin case to the self-shielded phase
occurs abruptly,
we also consider the emissivity of a \emph{condensed phase cut} case 
as the most conservative for the cooling radiation.
There we set the emissivity of the self-shielded gas particles to zero.

Hence, in the following analyses, we consider three possibilities.
First is the \emph{optically thin case} that assumes 
a spatially uniform UV background for every gas particle.
Second is the \emph{self-shielding corrected case} described above 
that uses an attenuated UV background for each gas particle 
appropriate for the local optical depth.
Third is the \emph{condensed phase cut case} where we set the emissivity of gas with 
${\rm log\ } T<4.5$ and $\rho/\bar{\rho}_b > 10^3$ to zero.
We emphasize again that while none of these possibilities are rigorously correct,
they range from the most optimistic (1) to the most conservative case (3).
Note again that case (1) is appropriate for \ion{He}{2}, 
but the full range of cases should be considered for \ion{H}{1}.

\section{RESULTS}

\begin{figure*}[t]
\epsscale{1.0}
\plotone{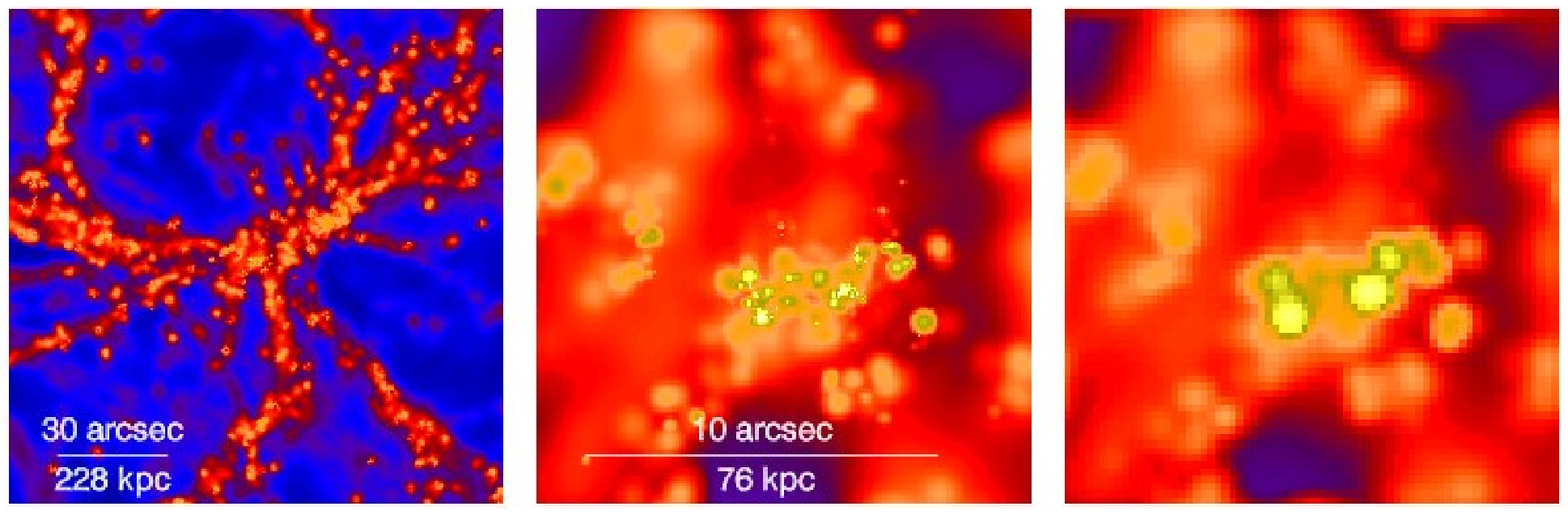}\\
\plotone{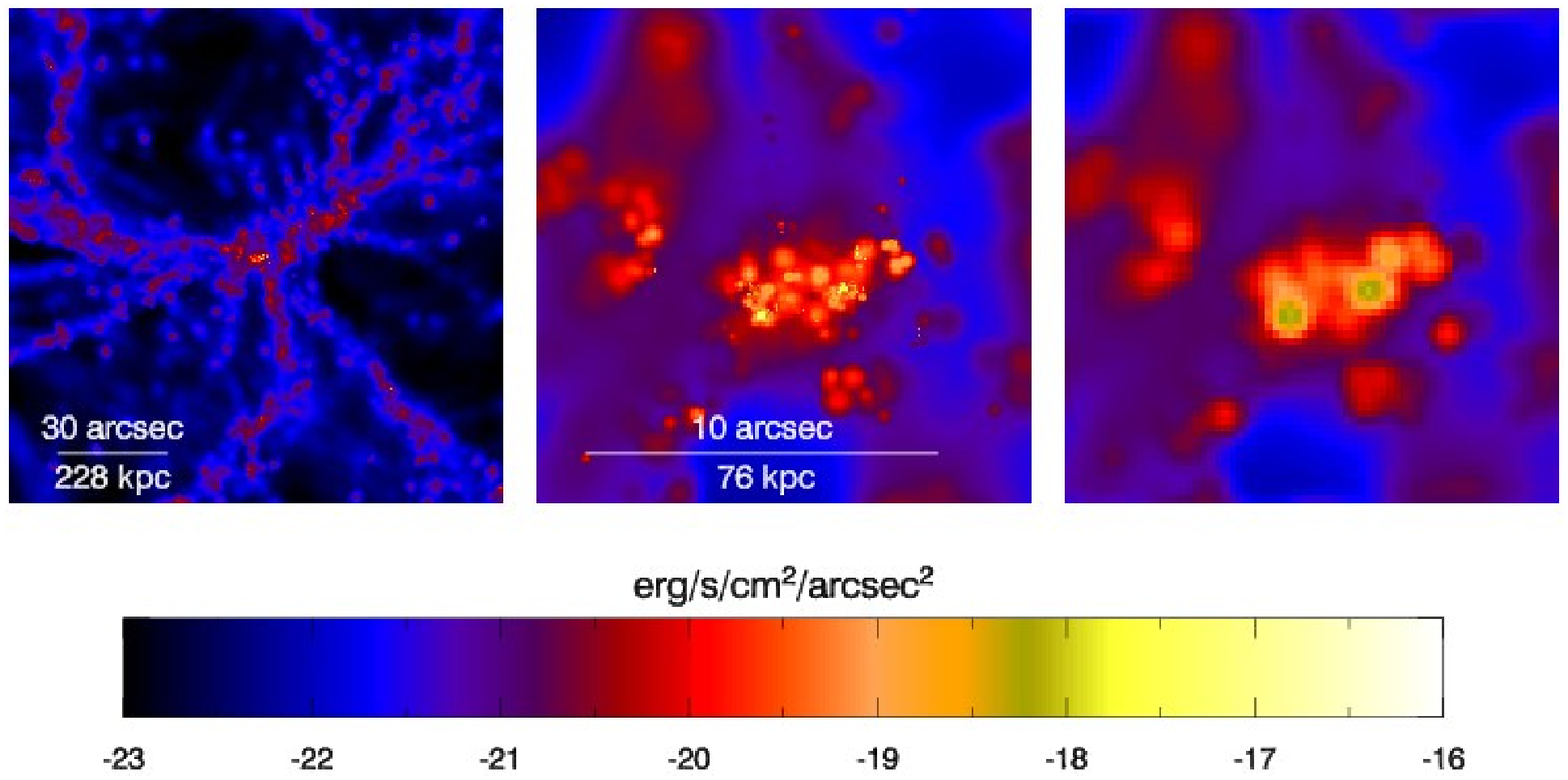}
\caption{
\Hlya\ (\emph{top}) and \Heha\ (\emph{bottom}) cooling maps for the 11 Mpc
simulation at $z=3$.  The line of sight depth is $\Delta z\simeq0.019$.
The left panels show a part ($\onequarter \times \onequarter$) of our simulation,
the middle panels show the brightest region at a finer pixel scale
($\sim 1.5\ h^{-1}\ $kpc per pixel), and
the right panels show the cooling maps convolved with a 0\farcs5 FWHM Gaussian filter
to mimic a typical ground-based observation (re-binned to 0\farcs2 per pixel).
Note that we include \lya\ and \Heii emission from the IGM
assuming that the emissivity of the condensed phase is zero,
the most conservative case.
The \lya\ cooling radiation from the gas around the forming galaxies
will be observed as a diffuse and extended blob above $\sim 10^{-18\ }\SB$,
the current flux limit of ground-based detections ($R=100$),
whereas \Heii will be almost point source-like at current detection limits.
}
\label{fig:maps:z3}
\end{figure*}

\subsection{Cooling Maps}
\label{sec:map}
We generate \hlya\ and \heha\ cooling maps at $z=$ 2 and 3
by applying our line emissivities 
to each pixel element and integrating them along the line of sight.
The temperature and density of each volume element is computed
using the usual SPH smoothing kernels, and the abundances of ionic species
are calculated from these smoothed quantities.
The thickness of the 11 Mpc simulation along the line of sight
is $\Delta z \simeq 0.013$ and $0.019$ for $z=2$ and $3$, respectively.
We convert these cooling maps into surface brightness maps 
using our adopted cosmology. 
Figure \ref{fig:maps:z3} shows the \hlya\ and \heha\
cooling maps for the 11 Mpc simulation at $z=3$.
We show the cooling maps for the condensed phase cut case
to represent the most conservative prediction.
The left panels show a part ($\onequarter \times \onequarter$)
of our simulation where the filamentary structure of the IGM 
--- the so-called ``cosmic web'' --- is evident.
In the middle panels, we show the brightest region
(also the most over-dense region for $z=3$) at a finer pixel scale
($\sim 1.5\ h^{-1}\ $kpc per pixel; 
half of the spatial resolution of our simulation).
The right panels show the cooling maps convolved 
with a 0\farcs5 FWHM Gaussian filter to mimic a typical ground-based observation.

As shown in Figure \ref{fig:maps:z3}, 
the \lya\ cooling emission from the IGM is somewhat extended 
above a surface brightness threshold of $10^{-18\ }\SB$
(the current limit of ground-based detections),
whereas the \heii emission will be seen almost as a point source. 
We will refer to these extended \lya\ cooling sources in our simulations 
as \lya\ blobs hereafter.
In the cooling maps, 
we include only \lya\ emission from the IGM, not from star formation 
(i.e. from photoionization caused by massive stars followed by recombination).
However, we find a compact group of stars and/or star-forming particles,
i.e., galaxies, at the center of each \lya\ blob.
Therefore, what we would actually observe are galaxies 
(or \lya\ emitters if dust absorption is negligible) embedded within the \lya\ blobs.

\begin{figure*}[t]
\plotone{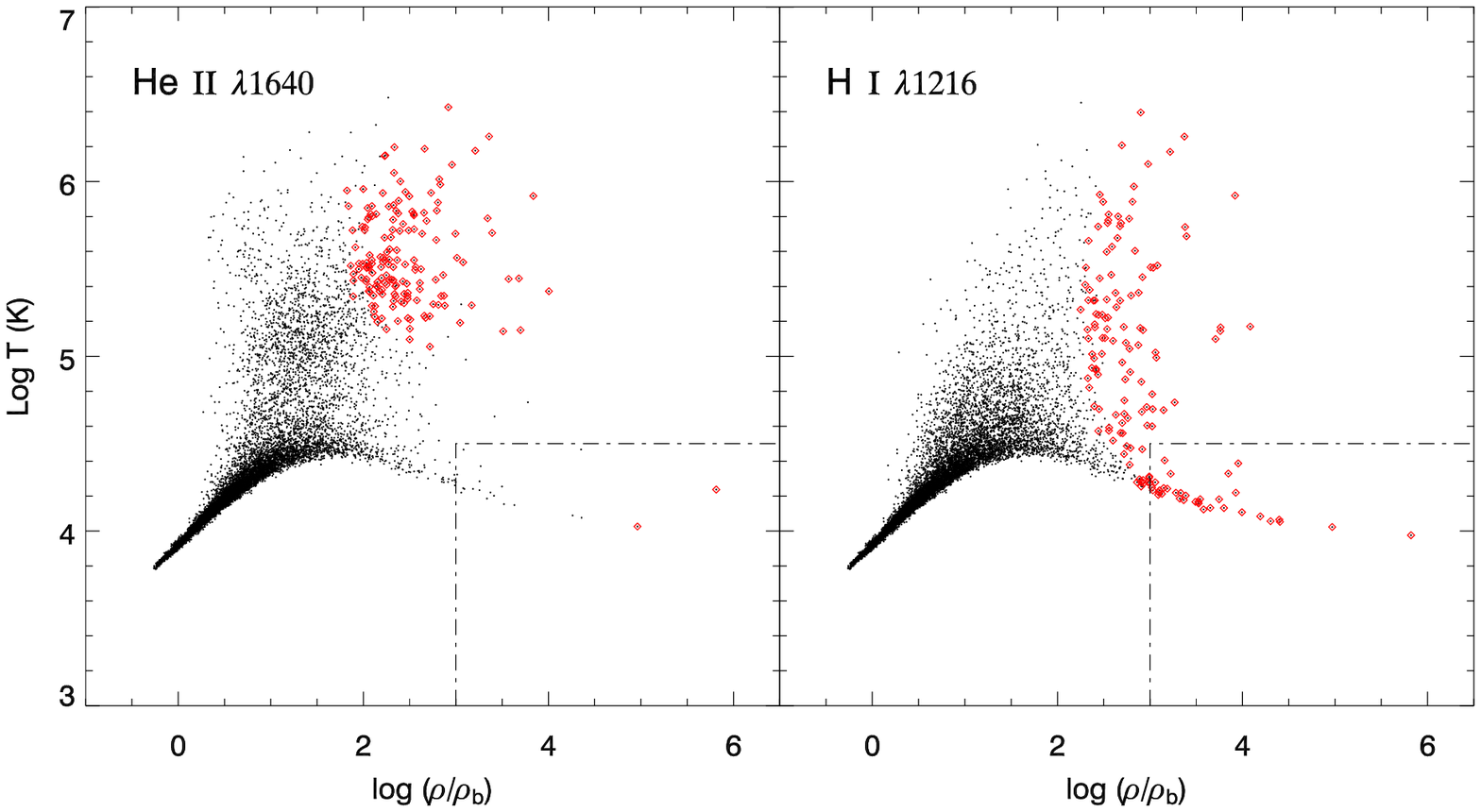}
\caption{
``Luminosity-weighted'' phase diagram for $100\times100$ lines of sight
through the simulation at $z=3$ for \Heha\ (\emph{left}) and \Hlya\ (\emph{right}).
The diamonds indicate the lines of sight
that have \Heha\ or \Hlya\ surface brightnesses larger than $10^{-19}\ \SB$.
The condensed phase of the IGM ($T<10^{4.5}$ and $\rho/\bar{\rho_b} > 10^3$)
is represented by the dot-dashed lines.
The brightest \Heha\ blobs occupy a specific range of
temperature ($10^5<T<10^6$) and density ($10^2<\rho/\bar{\rho_b}<10^3$),
while the brightest \Hlya\ blobs tend to come from the high density gas.
Thus the \Hlya\ cooling map is greatly affected by the condensed phase of the IGM.
To be conservative, we thus exclude the condensed phase gas particles
in generating the cooling maps (Fig. \ref{fig:maps:z3}).
}
\label{fig:phase}
\end{figure*}

The most uncertain factor in generating the cooling maps is 
how much the self-shielded gas contributes to the emission.
To quantify this factor globally in the simulations,
we consider the \hlya\ and \heha\ luminosity-weighted temperature and density plots 
for the optically thin (optimistic) case  in Figure \ref{fig:phase}.
To make these phase diagrams, we extract temperature and density profiles 
for 100$\times$100 evenly-spaced lines of sight, apply our emissivities to 
each radial bin, and integrate the temperatures and densities
with the \hlya\ and \heha\ luminosities as weighting factors.
Thus each diagram represents the phases that we could actually probe
by observing each line.
In each phase diagram, the condensed phase of the IGM is delineated 
by dot-dashed lines.
The sharp edge of the condensed phase at $T\sim10^4\ \K$ arises from
the lack of metal-line cooling in our simulations.
The diamonds in Figure \ref{fig:phase} indicate the lines of sight that have 
\heha\ or \hlya\ surface brightnesses larger than $10^{-19}\ \SB$. 
Most \heha\ emission comes from a specific range of temperature ($10^5<T<10^6$) 
and density ($10^2<\rho/\bar{\rho_b}<10^3$) that is remote from the condensed phase, 
whereas the brightest \hlya\ blobs have significant amounts of condensed phase gas.
Because self-shielding becomes important in the condensed phase, 
we exclude this phase in calculating the cooling maps (in Fig. \ref{fig:maps:z3})
to produce our most conservative predictions. 
As we expected, this cutoff does not affect the \heii cooling maps seriously, 
but does affect the \hlya\ cooling map dramatically, 
as gas particles with $T < 10^{4.5\ }\K$ cannot contribute 
to \heii collisional excitation cooling 
but only to \hi collisional excitation cooling.
Therefore, our predictions for the \heha\ cooling radiation 
are far more robust than for \hlya.

\begin{figure}
\epsscale{1.1}
\plotone{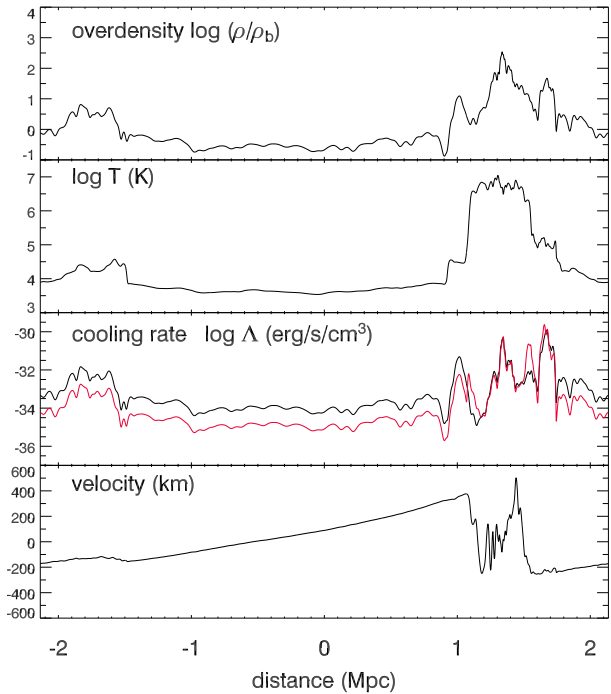}
\caption{
Profiles of density, temperature, \Hlya\ (\emph{bold}) and \Heha\ (\emph{light})
cooling rates, and velocity for a line of sight. Owing to the complicated structure
of the density and the turbulent velocity fields, it is difficult to
predict how much the radiative transfer blurs the surface brightness of
\lya\ cooling blobs. However, because the IGM is optically thin to He II,
our \Heii cooling maps (Fig. \ref{fig:maps:z3}) should be accurate.
}
\label{fig:los}
\end{figure}

The next factor that could modify the \lya\ cooling maps is 
the radiative transfer of \lya\ photons, 
which could alter the shapes and surface brightness profiles 
of the blobs substantially.
Both the \hlya\ and \helya\ photons produced in the optically thick medium 
will be transported to the outer region by resonant scattering 
until the optical depth becomes smaller than $\tau \sim 2/3$.
\lya\ photons escape eventually by scattering into the optically thin damping wing 
in the frequency domain, unless they are extinguished by dust. 
We expect that the IGM at $z\sim3$ contains little dust and that 
the cooling emission from the IGM is sufficiently far from the star-forming regions
since we exclude the high density gas particles in our condensed phase cut case.
Therefore, the net effect of the resonant scattering 
in the spatial and frequency domains is to smooth the surface brightness
out to the last scattering surfaces.
For example, \citet{Fardal} resorted to resonant scattering 
to explain the observed size of the Steidel blobs.
However, owing to the complicated structure of the density and 
to turbulent velocity fields,
it is difficult to predict how much radiative transfer blurs
the surface brightness of the cooling blob.
The large bulk motions will especially affect the transfer of \lya\ photons.
Depending on the optical depth and velocity field, 
photons can often undergo very little spatial diffusion and just random walk 
in velocity space until they reach a frequency where the optical depth 
is $\sim 1$ \citep{Zheng}. 
A Monte Carlo \lya\ radiative transfer calculation would be an ideal tool
to make more realistic spatial and frequency maps of \lya\ cooling radiation 
\citep{Zheng,Kollmeier}.
%
%
In Figure \ref{fig:los}, 
we show the profiles of density, temperature, velocity, and
\ion{H}{1} and \heii emissivity for a line of sight 
toward a typical \lya\ blob to illustrate 
the complicated structure of these quantities.
Because we do \emph{not} take into account these radiative transfer effects 
in our cooling maps, the \hlya\ cooling map 
(the upper middle panel in Figure \ref{fig:maps:z3})
should be smoothed to better represent reality.
In contrast, 
because most \heii resides in the ground state, 
making the IGM optically thin to the \heha\ line, 
our \heii cooling maps should be accurate.

Besides gravitational cooling and photoionization caused by the UV background, 
supernova feedback is another energy source included in our simulations.
When stars form in the simulations, supernova feedback
energy is deposited into the surrounding gas in the form of heat.
Thermal energy deposited into dense, rapidly cooling gas is
quickly radiated away, so feedback contributes somewhat
to the cooling emission.
However, we find that our density-temperature cutoff 
for the condensed phase effectively removes all the star-forming gas 
particles at each time step. Because the supernova thermal input is 
directly proportional to the star formation rate,
our cooling maps without the condensed phase should not
be seriously contaminated by supernova feedback energy.
Thus the cooling maps in Figure \ref{fig:maps:z3} 
with the condensed phase removed
are still robust lower limits of the flux from the gravitational cooling.
\citet{Fardal} also show that while the re-radiated supernova energy
dominates at lower luminosity, gravitational cooling becomes the dominant 
source as the mass and luminosity increase.

\subsection{Properties of Cooling Sources}
\label{sec:properties}
To study the properties of individual \hlya\ and \heha\ sources like the ones
shown in the cooling maps, we identify discrete groups of gas particles 
associated with individual dark matter halos and then calculate the total \lya\
and \heha\ luminosities for these sources.
To find the dark matter halos, we apply a friends-of-friends algorithm 
with a linking length that is 0.25 times the mean inter-particle separation.
We count a gas particle as a member of the source, i.e. blob, associated 
with the dark matter halo if the distance from the potential center is 
less than the virial radius of the halo.
We then add the \lya\ and \heha\ luminosities of the particles 
to obtain the total cooling luminosities of the blob.
We restrict our analysis to blobs with more than 64 gas particles
and 64 dark matter particles to mitigate numerical resolution effects.
Thus the smallest halo in the 22 Mpc simulation has a gas mass of
$M_{\rm gas}  = 6.3\times10^{9} \msun$ and dark matter mass of
$M_{\rm dark} = 5.0\times10^{10} \msun$. 
These masses decrease to 
$M_{\rm gas} = 8.5\times10^8 \msun$ and 
$M_{\rm dark} = 6.8\times10^{9} \msun$ in the 11 Mpc simulation.

\begin{figure*}
\epsscale{0.8}
\plotone{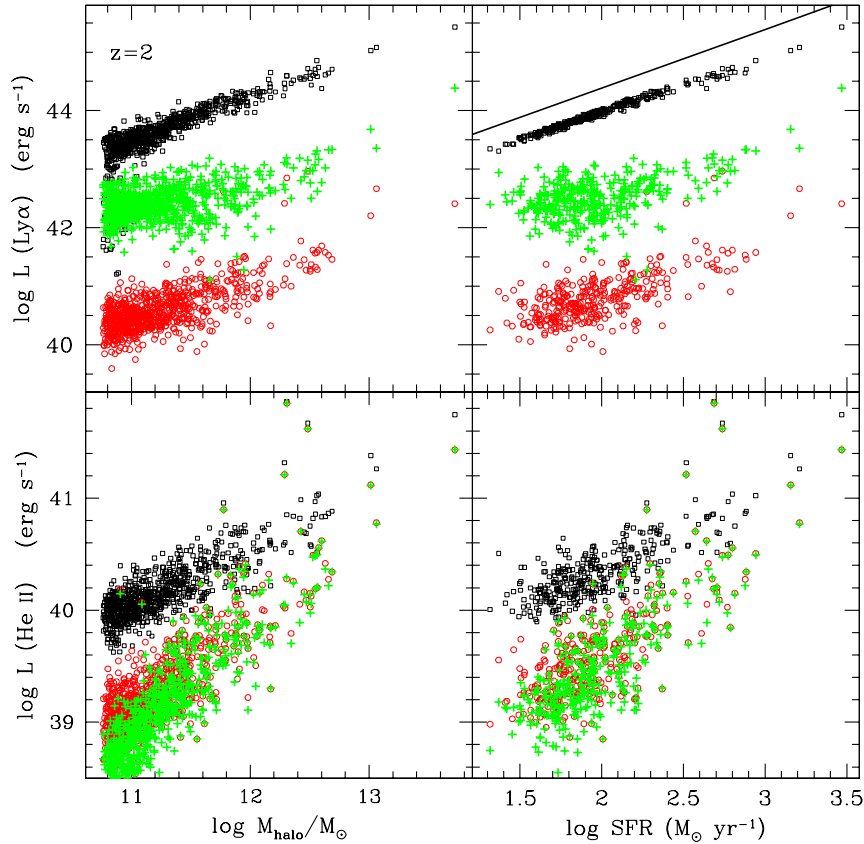}
\caption{
\lya\ and \Heii luminosity as a function of the halo viral mass and
the star formation rate in the 22 Mpc simulation at $z=2$.
The open squares, crosses and circles represent
the three different emissivity predictions:
the optically thin case, the self-shielding correction case, and
the condensed phase cut case, respectively.
In the right panels, we plot only blobs with a baryonic (star + gas) mass
larger than 200 $m_{SPH}$. Below this mass limit,
the derived star formation rates are not reliable owing to
our limited resolution.
As discussed in the text,
\lya\ luminosity changes dramatically depending on the prescription
used for the self-shielded phase. The correlations between the cooling luminosity,
halo mass, and SFR are as one generally expects:
the more massive a galaxy is, the more gas accretes onto the galaxy,
resulting in more cooling radiation and a higher star formation rate.
The solid line in the upper right panel represents the \lya\ emission
expected from star formation, assuming a conversion factor,
$f_{\rm Ly \alpha}=2.44 \times 10^{42\ } {\rm ergs\ s^{-1}}$,
for a $1 M_{\sun\ } {\rm yr^{-1}}$ star formation rate with no dust absorption,
no escaping ionizing photons, a Salpeter IMF, and solar metallicity.
Note that under these assumptions the \lya\ emission from star formation
always dominates the cooling emission from the surrounding IGM.
}
\label{fig:correlation}
\end{figure*}

\hlya\ and \heha\ cooling luminosities show tight correlations 
with halo mass and star formation rate (Fig. \ref{fig:correlation}).
The open squares, crosses and circles represent the luminosities for 
the three different emissivities discussed in \S \ref{sec:self}:
the optically thin case, 
the self-shielding correction case, and 
the condensed phase cut cases, respectively.
The correlations 
are as one would expect:
the more massive a galaxy is, the more gas accretes onto the galaxy,
resulting in more cooling radiation and a higher star formation rate.
The distribution of cooling luminosity is continuous, and
we do not find any evidence that extended \lya\ or \heii emission originates 
only from high-mass systems or high density regions.

The solid line in the upper panel in Figure \ref{fig:correlation} indicates 
the \lya\ emission due to recombination from stellar ionizing photons.
We assume a conversion factor of 
$f_{\rm Ly \alpha}=2.44 \times 10^{42\ } {\rm ergs\ s^{-1}}$ 
for a $1\ M_{\sun\ } {\rm yr^{-1}}$ star formation rate 
with no dust absorption, no escaping ionizing photons, a Salpeter IMF, and 
solar metallicity.
Under these assumptions
the \lya\ emission from star formation in galaxies always dominates 
the \lya\ emission from the surrounding IGM, 
even in the (most optimistic) optically thin case.
This result is consistent with the predictions of 
\citet{Fardal} and \citet{Furlanetto:05}.
\footnote{In contrast, we do not find the
trend of \citet{Fardal} in which \lya\ from cooling radiation 
dominates the \lya\ from star formation
in more massive systems.  We suspect that this difference 
is a consequence of including a
photoionizing background in the simulation analyzed here.}
The strong correlation between the \lya\ cooling rate 
in the optically thin case (squares) and the star formation rate results from 
the fact that the gas in the condensed phase tends to satisfy 
the star formation criteria of the simulation 
and is likely to form stars in the next time step.
In contrast, the \heii emission caused by star formation is quite uncertain 
because only extremely low metallicity ($Z<10^{-5}$) stars can emit 
the hard ionizing photons necessary to ionize \ion{He}{2}. 
However, star formation at $z=2-3$ is unlikely to be dominated by Population III
or extremely low metallicity stars. 
Unlike for \lya, the contribution of star formation to \heii must be
negligible. We discuss this point in more detail in \S \ref{sec:dis}.

\begin{figure*}[t]
\epsscale{0.8}
\plotone{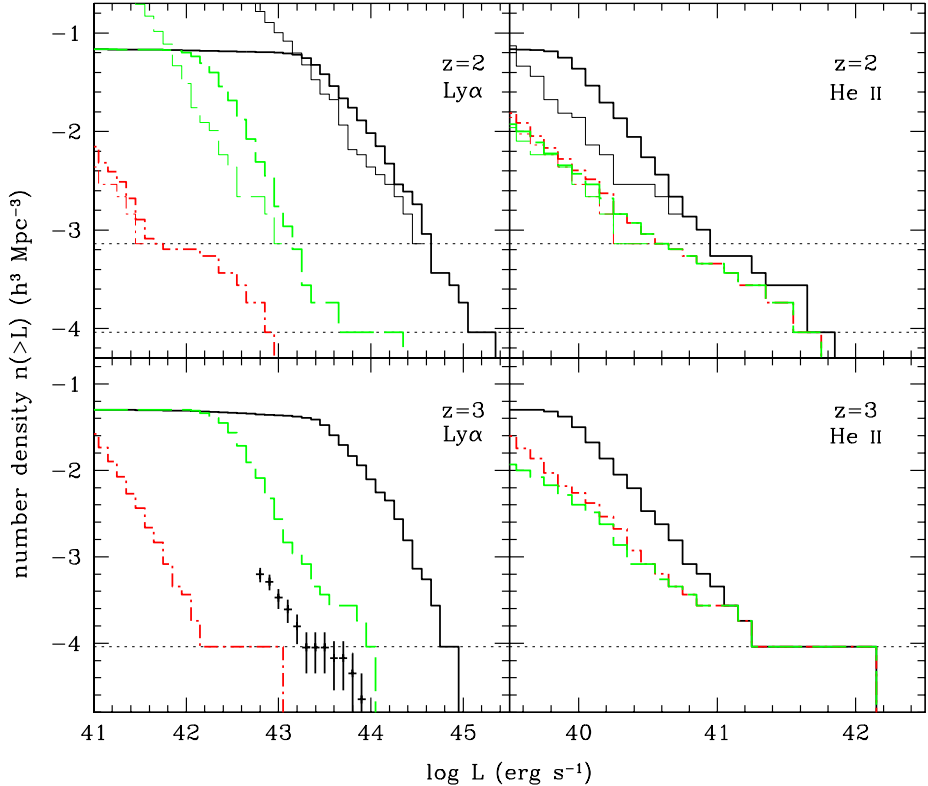}
\caption
{
Luminosity functions of \lya\ and \Heha\ cooling radiation
at $z=2$ (\emph{upper}) and $3$ (\emph{lower}).
The solid, dashed, and dot-dashed lines represent the optically thin,
the self-shielding correction, and the condensed phase cut cases, respectively.
In the upper panels,  for each emissivity case,
the LFs from the 11 and 22 Mpc simulations are denoted with light and bold lines,
respectively.
The horizontal dotted lines indicate the number density of one halo
in the 11 Mpc and 22 Mpc simulations.
Note that the 22 Mpc results extend to higher luminosity than the 11 Mpc results,
and that even the larger simulation may underestimate the density of
high luminosity systems.
In the lower panels, we show only results from the 22 Mpc simulation.
For comparison, we show the LF of the Subaru \lya\ blob sample
\citep[the crosses; ][]{Matsuda}.
The Subaru \lya\ blobs could arise from a variety of mechanisms (see text),
including cooling radiation.
It is suggestive, however, that their LF lies between our predictions
for the self-shielding correction and condensed phase cut cases.
Note also that
the \lya\ LF is very sensitive to the assumed emissivities,
whereas the \Heha\ cooling luminosity depends much less on the treatment of
the self-shielded or condensed phase of the IGM.
}
\label{fig:lf}
\end{figure*}

Figure \ref{fig:lf} shows the \lya\ and \heii luminosity functions (LFs) 
for the three emissivity cases at $z=2$ and 3. 
The LFs include emission only from the IGM.
The solid, dashed, and dot-dashed lines represent the optically thin,
the self-shielding correction, and the condensed phase cut cases, respectively.
The horizontal dotted lines indicate 
the number density of one halo in the simulation volume.
Note that the distributions extend to brighter blobs as we increase the 
simulation volume
because the larger simulations contain higher mass systems.
Note also that the \lya\ luminosity function is very sensitive 
to the assumed emissivities, 
whereas the \heha\ cooling luminosity depends much less 
on the treatment of the self-shielded or condensed phase of the IGM.
This large variation of the \lya\ luminosity is consistent 
with the results of \citet{Furlanetto:05}.

\subsection{Detectability and Observational Strategy}
\label{sec:detectability}

To estimate the detectability of cooling emission from the extended sources,
we convert the rest-frame cooling maps at $z=2$ and $3$ into observed
surface brightness maps and rebin them with a pixel scale of 
$0\farcs5\times0\farcs5$ 
to mimic the independent resolution elements of ground-based observations 
(Fig. \ref{fig:maps:z3}).
Figure \ref{fig:histogram} shows the surface brightness distributions 
of the rebinned cooling maps at $z=2$ and $3$,
assuming the conservative condensed phase cut case.
Note that the distributions depend strongly on 
the size of the bins in the surface brightness maps,
because the bright, small-scale structures are smoothed out by binning.
We express each surface brightness distribution in terms of the number of binned
pixels per comoving volume and also the number of pixels
per square arcmin if one were to observe through a $R=100$ narrow band filter.
The projected angular extents of the 11 Mpc simulation at $z=2$ and $3$ 
are $11.4'$ and $9.4'$, respectively.
The depth of the 11 Mpc simulation
is $\Delta z \simeq 0.013$ and $0.019$ for $z=2$ and $3$, respectively.

Deep, wide-field ($\sim30'\times30'$), narrow-band ($R>100$) imaging
is an effective way to detect cooling radiation,
because sky noise dominates in this low surface brightness range.
For example, the average sky background at 6500\AA\ on the ground is
$\sim10^{-17}$ $\mathrm{ergs\ s^{-1}\ cm^{-2}\ arcsec^{-2}\ {\AA}^{-1}}$, 
comparable to our estimates for the brightest blobs.
In Figure \ref{fig:histogram}, we show the $5\ \sigma$ detection limits 
for typical $R=100$ narrow band imaging
with an 8m-class telescope and for $R=1000$ imaging with a hypothetical 30m telescope.
We assume a peak system throughput of $\sim$ 35\%, 
a Mauna Kea sky background (for the 50\% dark condition), and a 30-hour exposure time.
We estimate the signal-to-noise ratios for one binned pixel 
($0\farcs5\times0\farcs5$), which corresponds to $\ga2\times2$ instrumental 
pixels in ground-based CCD detectors.

\subsubsection{{\rm \hlya}}
We predict that \hlya\ cooling emission from the brightest blobs at $z=2$ and $3$ 
is detectable by 6-8m class telescopes with moderate resolving power ($R=100$).
The limiting sensitivity of current surveys for high-$z$ \lya\ emitters is 
$\sim 10^{-18}$ $\mathrm{ergs\ s^{-1}\ cm^{-2}}$
\citep[e.g.,][and references therein]{Malhotra}.
It is encouraging that even our most conservative predictions suggest that 
the \lya\ blobs arising from gravitational cooling radiation are detectable 
with a reasonable amount of telescope time.

The \lya\ surface brightness of the largest system in our $z=3$ simulation
($M_{\rm halo} \sim 6.5 \times 10^{12} h^{-1} \msun$),
corresponding to the brightest blob in Figure \ref{fig:maps:z3}, 
is consistent with the \emph{mean} surface brightnesses of 
the \lya\ blobs of the \citet{Matsuda} sample
(represented with small vertical bars in the Figure \ref{fig:histogram}).
Note that our predicted \lya\ blob luminosities depend on 
the different emissivities for the self-shielded phase and/or 
the exact location of our density-temperature cut of the condensed phase.
Though the surface brightnesses of the predicted and observed blobs are consistent,
the \emph{luminosity} of our brightest \lya\ blob is fainter 
than that of observed blobs (see Fig. \ref{fig:lf}),
possibly because of our conservative assumptions for the self-shielded phase.

\subsubsection{{\rm \heha}}
A pixel-by-pixel comparison of the \hlya\ and \heha\ cooling maps reveals that, 
without the condensed phase, 
the \heha\ flux is always $\ga 10 \times$ fainter than that of \hlya.
The \heha\ emission could be $1000 \times$ fainter than \lya\ 
in the optically thin case, i.e., the most optimistic \lya\ prediction.
Nonetheless, detection of the \heha\ cooling line from $z=2$ sources is 
clearly feasible with 6-8 meter class telescopes, and even possible at $z=3$.
Though the number statistics of bright blobs are limited 
by the relatively small volume of our simulations,
we expect one source in the 11 Mpc simulation
and six sources in the 22 Mpc simulation at $z=2$ 
with areas of $\ga 0\farcs5\times0\farcs5$ above 
the surface brightness detection threshold of $\sim 5 \times 10^{-18}\ \SB$
($R=100$ arrow in Figure 8).
Thus the space density of the sources from which we could detect 
not only \lya\ but also \heha\ emission with narrow band imaging
corresponds to a comoving number density of 
$\sim 5-7\times10^{-4}\ h^3\ \mathrm{Mpc}^{-3}$ or 
$\sim 0.02\ \mathrm{arcmin}^{-2}$ per $R=100$ filter
($\Delta z=0.03$; $\Delta \lambda \simeq 49$\AA\ for \heha).

\begin{figure*}[t]
\epsscale{1.1}
\plottwo{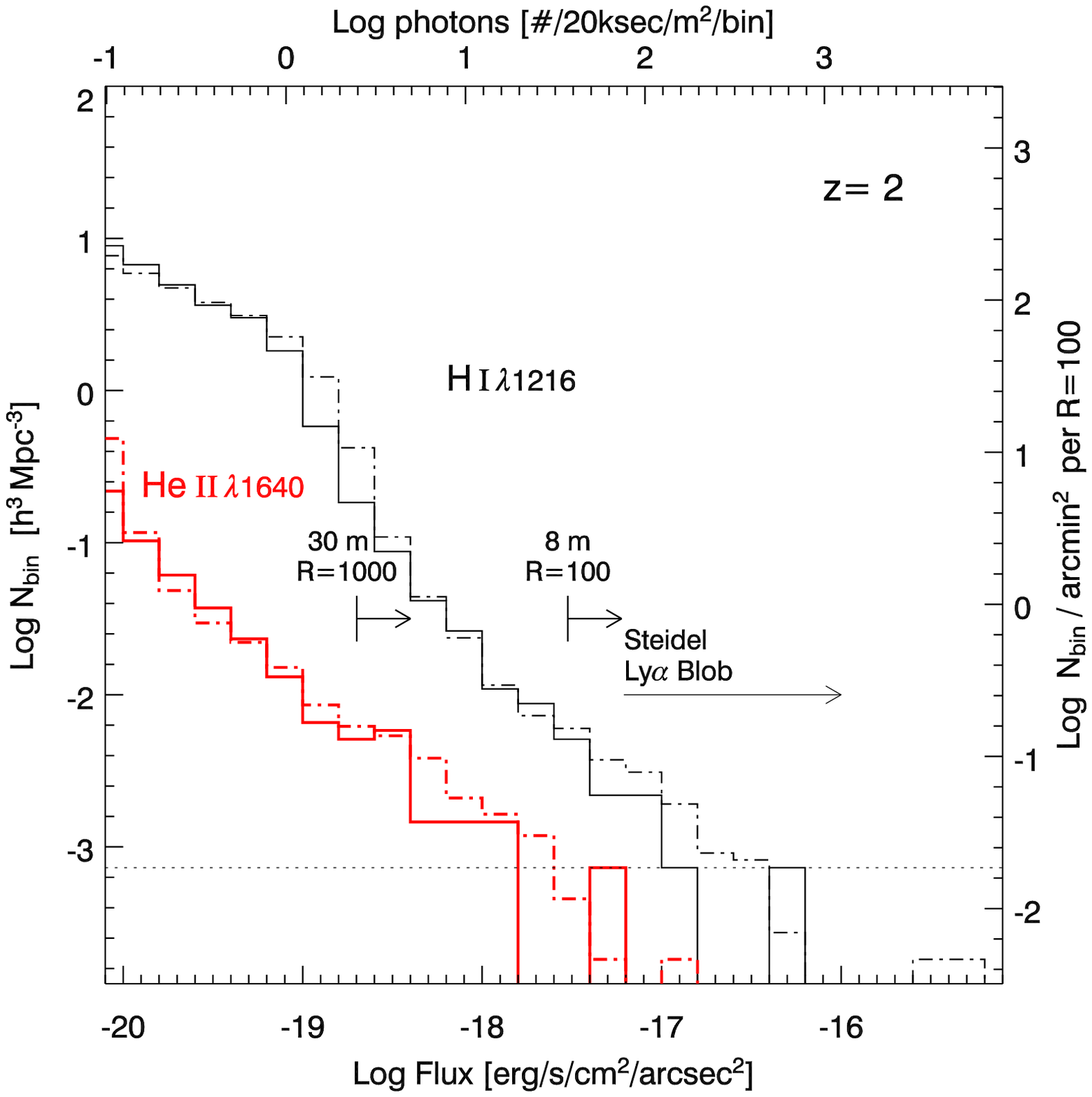}{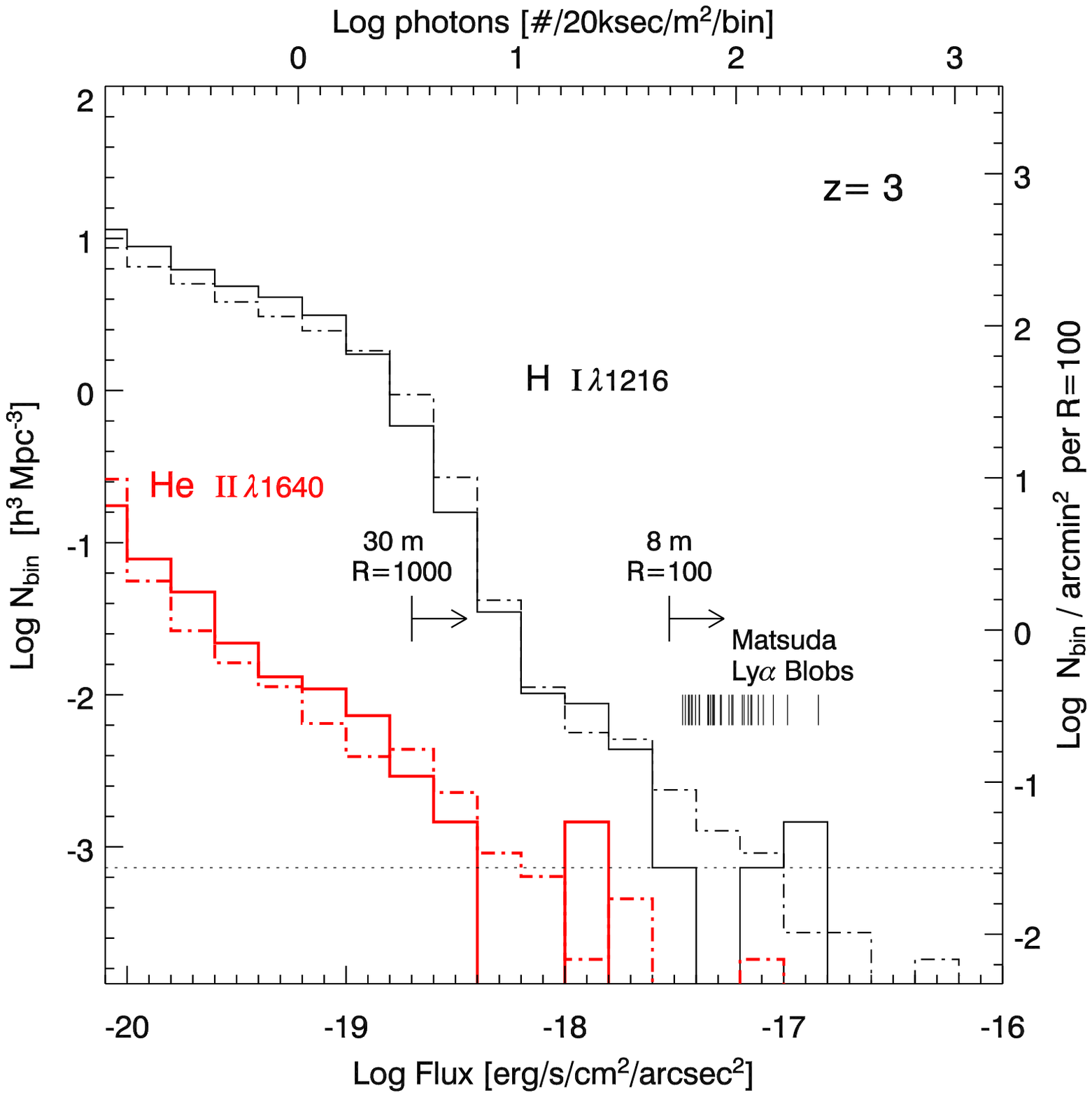}
\caption{
Distributions of surface brightnesses in the cooling maps for the IGM at $z=2$ and 3.
Bold and light solid lines represent the
surface brightness histograms of rebinned ($0\farcs5\times0\farcs5$) pixels
for \Heha\ and \Hlya, respectively, in the 11 Mpc simulation.
We show the $\mathrm{S/N}>5$ detection limits
for a 30-hour observation with an 8-meter telescope and $R=100$ narrow-band filter,
and with a 30-meter telescope and $R=1000$ filter.
The right $y$-axis represents the number of binned pixels
per square arcminute per the redshift width for the $R=100$ filter.
The dotted lines at the bottom of each panel indicate one detection in the simulation.
The dot-dashed lines represent the surface brightness distributions
for the 22 Mpc simulation.
Detection of the brightest \Hlya\ and \Heha\ blobs at $z=2$ and $3$ is possible with
deep narrow-band imaging on a 6-8 meter class telescope.
Note that
the larger simulation extends the bright tail of the surface brightness distribution.
Even larger simulation volumes might therefore predict still brighter,
thus more easily detectable, systems.
}
\label{fig:histogram}
\end{figure*}

If we consider the larger survey volume 
typically accessible by modern narrow band imagers,
we could expect better survey efficiency than that described above.
Because cosmological simulations do not contain power on scales larger
than their finite sizes, the largest objects in the simulation are 
typically underestimated in number and size.
For example, the luminosity functions and surface brightness distributions 
in Figs. \ref{fig:lf} and \ref{fig:histogram} extend their bright limits as 
the volume of the simulation increases.
Therefore, one might expect that the detection of even brighter blobs 
by surveying more volume.
For example, many current wide-field imagers and spectrographs have 
half degree field of views, so it is possible to survey 
a volume $\sim 17$ times larger than encompassed by our 11 Mpc simulation at $z=2$
(or a volume $\sim$ twice that of the 22 Mpc simulation).
Therefore, if we na\"ively extrapolate the number of detectable \heii sources 
in our 11 Mpc and 22 Mpc simulations, then
17 ($\pm17$) and 13 ($\pm5$) \heii sources would be detected, respectively.
The numbers within parentheses indicate Poisson errors.

Our results predict that bright \heii sources are always bright \lya\ cooling blobs.
Observationally, the difficulties in searching for \heii cooling sources 
could be eased 
(1) by pursuing narrow-band \lya\ imaging first, detecting \lya\ blobs, 
and looking for \heha\ emission in those blobs with follow-up observations, or 
(2) by adopting a combined multislit spectroscopy + narrow-band filter approach 
\citep{Martin,Tran} to identify \lya\ blobs 
(which can then be targeted for \ion{He}{2}).

The latter technique is potentially quite effective 
despite the faint surface brightness of the \lya\ and \heii blobs.
This technique employs multiple parallel long slits with a narrow band filter 
that limits the observed spectral range to a few hundred angstroms and, 
by dispersing the sky background, 
achieves better sensitivity than narrow-band imaging alone.
In the sense that this technique trades off survey volume (or sky coverage)
to go deeper in flux, it is about as
efficient as simple narrow band imaging
for surveys of \lya\ emitters, which are not as extended as \lya\ blobs.
However, the multi-slit window technique is superior to narrow band imaging for 
low surface brightness objects like the \lya\ and \heii blobs discussed here.
It has the further advantages that
1) it provides the spectral and kinematic data necessary to distinguish the origins 
of blob emission (\S \ref{sec:dis}) and 
2) it enables us to exclude contaminating emission lines,
such as H$\alpha$, H$\beta$, [\ion{O}{3}], and [\ion{O}{2}], 
from nearby star-forming galaxies by measuring the line shapes 
(e.g. line asymmetries and line doublets) and blueward continuum.
If this technique is employed with large field-of-view imagers,
the survey volume covered by the multiple slits is still reasonably large 
($\sim 10\%$ of the whole field of view).

It is possible to search for \heii cooling radiation 
at lower redshifts than $z\sim2-3$.
For example, $z\approx1.5$ is the lowest redshift at which \heha\ still lies 
at an optical wavelength ($\lambda_{obs}\approx4100\ $\AA).
Because metal abundances in the IGM do not change very much over $z \sim 2-4$ 
\citep{Schaye}, 
it is unlikely that our basic assumption of a primordial composition 
(\S \ref{sec:sim}) is violated seriously at $z\approx1.5$. 
Any blind search for \heha\ blobs at $z=1.5-2$ will be
contaminated by \hi \lya\ emission from $z\approx3$ sources
if only one emission line is identified in the spectrum.
In this case, the blob's redshift could be further constrained 
by obtaining a redshift for the galaxy it surrounds.

\heha\ cooling radiation at very low redshifts ($z\la0.5$)
is potentially detectable in the ultraviolet using UV satellites or \emph{HST}.
For example, \citet{Furlanetto:03a} show that 
detection of the bright cores of \hlya\ emission from $z\la0.5$ sources
is feasible with deep wide-field UV imaging,
e.g. with \emph{The Galaxy Evolution Explorer} (GALEX) or 
the 
proposed \emph{Space Ultraviolet-Visible Observatory} \citep[SUVO;][]{Shull}.
Because \hlya\ and \heha\ trace different phases of the gas, as shown 
in Figure \ref{fig:phase},
combined observations of these two lines,
e.g., of their morphologies and line ratios,
would probe different phases of the IGM.

\subsubsection{\rm{\helya}}
In contrast to \hlya\ and \heha, \helya\ photons 
redshifted to wavelengths shorter 
than the photoionization edge of \hi and \hei (912\AA\ and 504\AA, respectively)
can be absorbed by neutral hydrogen and neutral helium.
Even if they escape the blobs, \helya\ photons are removed from the line of sight 
owing to cumulative absorption by the intervening neutral IGM,  
including the \lya\ forest and damped \lya\ systems.
We estimate the transmission of \helya\ through the intervening IGM
using Monte Carlo simulations as described in \citet{Moller} with 
the updated statistics of Lyman forest and Lyman limit systems
(i.e., number density evolution and column density distribution) from \citet{Jakobsen}.
For the emitters at $z=3$, we find that the transmission factor averaged over 
all lines of sight is $\sim$ 12\% and that 67\% (76\%) of the sightlines will have
transmission lower than 1\% (10\%).
Therefore, though the \helya\ emissivity is roughly $10\times$ higher than 
that of \heha\ (Fig. \ref{fig:cooling} in \S \ref{sec:CoolingCurves}),
we expect the \helya\ cooling map to be fainter than that of \heha\ in most cases
and to vary strongly from sightline to sightline.
\footnote{If one sightline does not have any Lyman limit or damped \lya\ systems,
it will have $\sim$87\% transmission on average due only to Lyman forest systems.} 

\helya\ photons can also be destroyed by the \hi and \hei inside the blobs,
because \helya\ photons experience a large number of scatterings before escaping. 
The destruction probability by \hi and \hei atoms per scattering is given by
\begin{equation} 
\epsilon = \frac{n_{\mathrm{H\sss I}} \sigma_{\mathrm{H\sss I}} + n_{\mathrm{He\sss I}} \sigma_{\mathrm{He\sss I}}}
                {n_{\mathrm{H\sss I}} \sigma_{\mathrm{H\sss I}} + n_{\mathrm{He\sss I}} \sigma_{\mathrm{He\sss I}} + n_{\mathrm{He\sss II}} \sigma_{\mathrm{Ly\alpha}}},
\end{equation}
where $\sigma_{\mathrm{H\sss I}}$ and $\sigma_{\mathrm{He\sss I}}$ are 
the photoionization cross sections of \hi and \hei at 304 \AA, respectively, 
and $\sigma_{\mathrm{Ly\alpha}}$ is the integrated scattering cross section of \helya.
The abundances of \hi and \hei atoms are smaller than for \ion{He}{2}, 
and their photoionization cross sections are also much smaller than the resonant
cross section of \helya\ by a factor of $\la 2\times10^{-5}$. 
We estimate that the destruction probability is $\epsilon \ga 5\times10^{-8}$ 
at a temperature of $T\sim10^{4.8}$ without applying the self-shielding correction.
In the self-shielded regions where more neutral hydrogen can reside,
this probability rises.
Thus the escape probability of a \helya\ photon from a blob is
$f_{\mathrm{IGM}} \simeq (1-\epsilon)^{N_{\tau}}$, 
where $N_{\tau} \simeq \tau^2$ is the number of scatterings 
required to escape the blob, 
if it is approximated by an optically thick slab.
For example, we obtain $f_{\mathrm{IGM}} \simeq 0.7\%$ 
for $\tau_{\mathrm{He\sss II}}=10^4$.
Therefore, we cannot ignore the absorption by \hi and \hei atoms 
in the high density gas.
However, the number of scatterings $N_{\tau}$ is very difficult to estimate correctly 
because of the complex density and velocity structure of the blob, 
unless one carries out full 3-D hydro-radiative transfer calculations
(which are beyond the scope of this paper). 
For certain geometries and velocity fields, 
bulk motions of the gas will help the \helya\ photons 
escape the blob with fewer scatterings \citep[e.g.,][]{Zheng}.


Owing to intervening absorption and the destruction inside
the blobs, \helya\ is the most uncertain cooling line we consider.
Although \helya\ is diminished significantly by the intervening IGM,
if the escape fraction from the IGM is significant ($f_{\mathrm{IGM}} \simeq 1$), 
the detection of \helya\  may not be out of question 
with a large aperture UV/optical optimized space telescope 
\citep[e.g., SUVO;][]{Shull}.
An advantage of observing \helya\ in the far ultraviolet in space is 
that the sky background is very low 
($\sim 10^{-23}$ $\SB$ $\mathrm{\AA^{-1}}$ at 1250 \AA) 
compared to the optical ($\sim 10^{-18}$ $\SB$ $\mathrm{\AA^{-1}}$ at 6500 \AA),
except for the geocoronal emission lines 
(e.g., \lya\ 1216 \AA~ and \ion{O}{1} 1304 \AA).
Thus, if these geocoronal emission lines could be eliminated 
with blocking filters or by adopting an L2 orbit,
the direct detection of \helya\ is feasible.
In this case, the detector noise --- especially the dark current --- 
will dominate.
Recent developments in UV detector technology are very promising, 
so the possibility of studying these blobs at those wavelengths remains open.

\section{DISCUSSION}
\label{sec:dis}
Until now, we have only considered the cooling radiation from gas
that is losing its gravitational energy,  
falling into a galaxy-sized dark halo, 
and ultimately forming stars.
Photoionization by these stars is another possible heating source for the blobs.
Starburst-driven superwinds or AGNs, which are not included in our simulations, 
are other potential blob energy sources 
\citep[e.g., see the discussions in][]{Steidel,Matsuda}.
Although the radiation from gas heated by these feedback
processes is not generally termed ``cooling radiation'',
the energy injected into the surrounding gas can also be released 
through line emission. Thus our estimates for cooling emission 
might be lower limits for the actual fluxes in the cooling lines.
In this section, we assess whether other \hlya\ and \heha\ sources 
overwhelm our gravitational cooling signals and then discuss 
how to discriminate among these other possible mechanisms
in order to use \hlya\ and \heha\ cooling lines to study gas 
infall into galaxies.

\subsection{Photoionization by Stellar Populations}
UV photons from massive stars in a galaxy or blob 
ionize the surrounding interstellar medium, and 
the recombination lines from these nebulae 
could contribute to the \lya\ and \heii line fluxes.
Generally, the recombination line luminosity is proportional to 
the star formation rate (SFR) and is given by
\begin{equation}
\label{eq:line}
	L_{\mathrm{line}} = e^{-\tau_{\mathrm{dust}}} 
	(1-f_{\mathrm{esc}})  f_{\mathrm{IGM}} f_{\mathrm{line}} 
	\left(\frac{\mathrm{SFR}}{\msun \mathrm{yr}^{-1}}\right),
\end{equation}
where $\tau_{\mathrm{dust}}$ is the dust optical depth for the ionizing continuum 
in the interstellar medium (ISM), $f_{\mathrm{esc}}$ is the fraction of 
ionizing photons that escape the star-forming galaxy, 
$f_{\mathrm{IGM}}$ is the fraction of photons that escape the surrounding IGM,
and $f_{\mathrm{line}}$ is the conversion factor
from the SFR to the line luminosity in $\mathrm{ergs\ s^{-1}}$.
This SFR conversion factor depends on the metallicity, 
initial mass function (IMF), and evolutionary history of the stars in the blobs
(e.g., a lower metallicity and a top heavy IMF produce more ionizing photons
and thus more recombination line photons).
\footnote{
Note that the equation (\ref{eq:line}) does \emph{not} take into account 
the dust absorption of \lya\ photons by ISM 
after \lya\ photons escape from the \hii region.
When the \hi optical depth in the ISM is high enough, 
\lya\ photon will experience a large number of resonant scatterings 
before escaping the galaxy. 
These scatterings will increase the effective dust optical depth and destroy
\lya\ photons preferentially.
However, the actual optical depth is likely to strongly depend on
the kinematics of neutral gas and the geometry of the galaxy,
which are hard to quantify.
}

The conversion factor for \hlya, $f_{\mathrm{1216}}$, 
is large enough to make it difficult to distinguish 
the cooling lines (of IGM origin) 
from the recombination-induced lines (of ISM origin).
For example, \citet{Schaerer} finds 
$f_{\mathrm{1216}} = 2.44 \times 10^{42}\ \mathrm{ergs\ s^{-1}}$
for a constant star formation history, solar metallicity, 
and a Salpeter IMF with a mass range of $1 - 100\ \msun$.
Thus for a $\mathrm{SFR}=10\ {\msun\ \mathrm{yr}^{-1}}$, 
$f_{\mathrm{esc}} \simeq 0.1$, $e^{-\tau_{\mathrm{dust}}}\simeq0.1$, 
and $f_{\mathrm{IGM}}\simeq1$, 
we obtain the observed flux, 
$F_{1216} \simeq 2.9\times10^{-17}\ \mathrm{ergs\ cm^{-2}\ s^{-1}}$,
due to the stars in a \lya\ blob at $z=3$,
which is comparable to the surface brightness of the brightest blobs 
in our simulations (see \S\ref{sec:detectability} for the blob luminosity functions).
Therefore, the contamination of the \hlya\ line by stars is not negligible,
unless the \lya\ photons from star-forming regions are heavily absorbed by a dusty ISM
(e.g., in highly obscured sub-millimeter galaxies 
or Lyman break galaxies with the damped \lya\ absorption).
Because \lya\ cooling radiation is produced sufficiently far from 
the star-forming region and thus should be less susceptible 
to dust attenuation than the \lya\ emission from the stellar populations,
it might be possible to isolate extended \lya\ cooling radiation
in these galaxies. 
However, in the case that \lya\ photons emitted by stars escape the galaxy
(e.g., \lya\ emitters),
it will be challenging to distinguish the \lya\ cooling radiation from 
the \lya\ produced by the stellar populations unless the various parameters 
such as $f_{\mathrm{esc}}$, $f_{\mathrm{IGM}}$, and SFR are fully constrained.

In contrast, 
\heha\ emission appears to be limited to very small metallicities 
($\mathrm{log}(Z/Z_{\sun}) \la -5.3$) and Population III objects,
because stars of solar or subsolar metallicities
emit few if any \heii ionizing photons \citep{Bromm,Schaerer,Tumlinson:03}.
Using \heha\ to detect the first hard-ionizing sources such as
metal-free stellar populations, the first miniquasars, 
or even stellar populations before the reionization epoch has been proposed 
\citep[e.g.,][]{Tumlinson:01,Oh,Barton}.
In this paper, we take advantage of this fact to discount the contributions
of stellar populations to the \heha\ cooling line.
Even for $Z=10^{-5}$, 
$f_{\mathrm{1640}} = 1.82 \times 10^{40}\ \mathrm{ergs\ s^{-1}}$
($\simeq 6\times10^{-4} f_{\mathrm{1216}}$) for 
an extremely top heavy IMF containing only stars in the range $50-500\ \msun$.
For the same assumptions used in the \lya\ calculation above,
we obtain an observed flux of 
$F_{1640} \simeq 2.2\times10^{-19}\ \mathrm{ergs\ cm^{-2}\ s^{-1}}$ 
from the blob stars at $z=3$, 
an order of magnitude below the brightest \heha\ blobs in our simulations.
Therefore, it is very unlikely that the \heha\ photons originating from 
stars contaminate the gravitational cooling emission,
unless significant numbers of metal-free stellar populations are forming.
Thus \heha\ cooling radiation is much less contaminated than \hlya\ 
by recombination lines originating from star-forming galaxies.

\begin{figure}
\epsscale{1.2}
\plotone{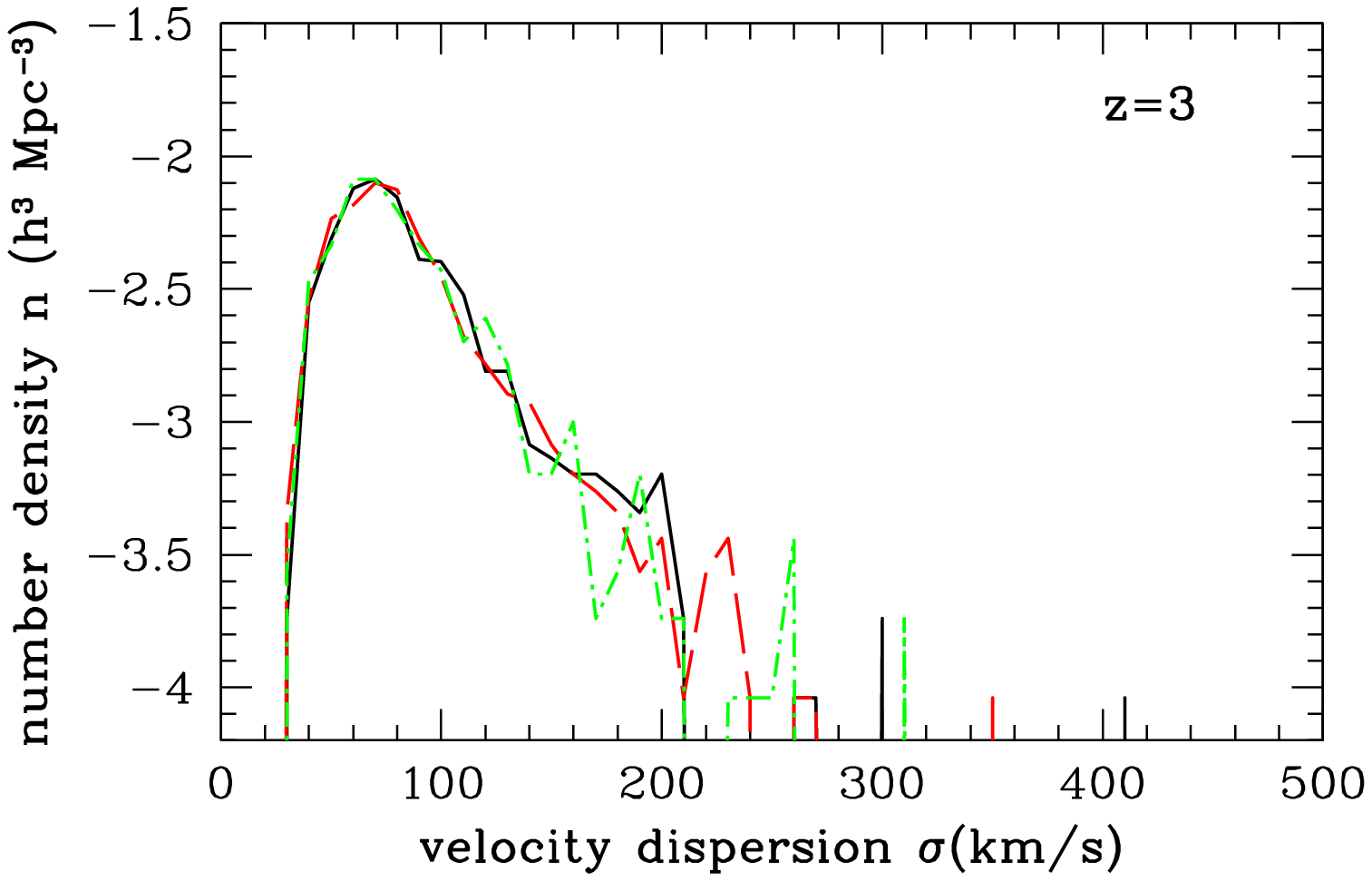}
\caption{
Distribution of \Heha\ flux-weighted velocity dispersion of the gas particles
associated with individual dark matter halos.
The different lines represent the velocity dispersions
in the $x$, $y$, and $z$ directions.
Note that the velocity dispersion due to gas accretion is less than 400 km s$^{-1}$,
in contrast to the typical galactic wind speed of
$\sim$ 400-800 km s$^{-1}$ \citep{Heckman}.
Thus the width of the optically thin \Heha\ line could be used as a diagnostic
to discriminate between the galactic wind and the gravitational cooling hypotheses
for powering \lya\ blobs.
}
\label{fig:veldist}
\end{figure}

The only caveat is 
the possibility of \heha\ emission arising not from stars directly, 
but from the hot, dense stellar winds of Wolf-Rayet (W-R) stars,
the descendants of O stars with masses of $M > 20-30\ \msun$.
W-R populations formed in an instantaneous starburst at high redshifts
would not seriously contaminate the \heha\ cooling radiation, 
because W-R stars are very short-lived ($\la$ 3 Myr) and their number relative 
to O stars (W-R/O) drops as the metallicity decreases below solar.
In the case of continuous star formation, 
a stellar population synthesis model \citep[Starburst99;][]{Leitherer} predicts 
that the maximum number of W-R stars is reached $\sim$10 Myr after the initial burst.
Using the \heha\ luminosity of a W-R star \citep{Schaerer:98} and
the number evolution of W-R stars from Starburst99 
under the assumptions of a Salpeter IMF ($1-100\ \msun$),
sub-solar metallicity ($Z \leqslant 0.4 Z_{\sun}$), and 
a massive SFR of 100 $\msun$ yr$^{-1}$ over at least 10 Myr,
we estimate the \heha\ line luminosity due to W-R stellar winds 
to be $\lesssim 10^{42}$ $\rm ergs\ s^{-1}$.
Although this \heha\ luminosity is comparable 
to the predicted \heii cooling radiation,
it is possible to discriminate between the two \heha\ sources in individual objects
because the emission from W-R winds should be much broader 
(e.g. $\sim$ 1000 km s$^{-1}$; see Fig. \ref{fig:veldist} in \S \ref{sec:superwind},
which is relevant here even though it is presented in the context of 
the galactic superwind scenario).

One way to test our predictions in this section is to look more closely 
at the \heha\ emission associated with high redshift star-forming galaxies,
i.e., the Lyman break galaxies (LBGs) with vigorous star formation rates.
\citet{Shapley} show that composite spectra of LBGs have
very broad ($\mathrm{FWHM} \sim 1500\ \mathrm{km\ s^{-1}}$)
\heha\ profiles regardless of their \lya\ emission strength.
While they attribute the \heii features to W-R stellar winds,
those authors have difficulty reproducing the strength of the \heii lines
using stellar population synthesis models with reasonable parameters.
Because of this inconsistency, 
we speculate that some fraction of the \heii features may
come from the cooling of gas falling into these galaxies along line of sight.
It would be worthwhile to obtain high signal-to-noise spectra of individual LBG's
and their surroundings to see if the \heii line is present,
especially outside the galaxy, and relatively narrow.

\subsection{Photoionization by AGNs}
AGNs inside the star-forming regions of blobs could photoionize the surrounding
gas and generate \heha\ as well as \hlya\ emission.
The predicted size (a few arcseconds) and surface brightness 
($\sim10^{-18}-10^{-16}\ \SB$) of an extended \lya\ blob enshrouding a quasar 
are consistent with the observed quantities \citep{Haiman:2001}.

How many AGN-powered sources might we expect in a \lya/\heii blob survey?
Unfortunately, it appears that there is no easy way 
to predict \lya/\heii luminosity from the surrounding IGM, 
because we do not know 
how much neutral IGM is distributed around the AGN. 
Therefore we take a conservative approach to estimate
the number of the AGN-powered sources. 
First, we establish a simplistic relationship 
between the induced \lya\ (or \ion{He}{2}) blob luminosity 
and the X-ray luminosity of the AGN, 
then we estimate the number of AGN-powered blobs
based on the known hard X-ray luminosity function of AGN at $z=2-3$.
If there are not many relative to the number of blobs powered 
by gravitational cooling radiation, then we could conclude that 
they are unlikely to complicate the interpretation of extended 
\lya/\heii sources.

We assume that all the ionizing photons from an AGN with 
a simple power-law spectrum, $L_{\nu} = L_0 (\nu/\nu_0)^{\alpha}$,
are absorbed by the surrounding medium and re-emitted as recombination lines.
The line (\lya\ or \ion{He}{2}) luminosity of the surrounding blob is then given by:
\begin{equation}
L_{\rm line} = c_{\rm line} Q = c_{\rm line}
  \int^{\infty}_{\nu_{LL}}\frac{L_0}{h \nu}\left(\frac{\nu}{\nu_0}\right)^{\alpha}, 
\end{equation}
where $Q$ is the number of ionizing photons emitted per unit time, 
$\nu_{LL}$ is the frequency of the Lyman limits for the hydrogen and \heii
($h \nu_{LL} = 13.6\ \eV$ and $54.4\ \eV$, respectively), 
and the line emission coefficient $c_{\rm line}$ in $\mathrm{ergs}$ is the energy of 
the line photon emitted for each \hi or \heii ionizing photon.
For case-B recombination with an electron temperature of $T_e = 30,000 \K$ 
and an electron number density of $n_e = 100\ \mathrm{cm^{-3}}$, 
we obtain $c_{1216} = 1.04\times10^{-11}$ and 
$c_{1640}=5.67\times10^{-12}\ \mathrm{ergs}$ \citep[c.f.][]{Schaerer}.
We adopt a spectral index $\alpha = -1.8$ for the extreme UV \citep{Telfer} 
and assume that this $\alpha$ is valid even in the X-ray.
The hard X-ray luminosity of AGN ($L_X$) is simply given by the integration of 
$L_{\nu}$ between 2 keV and 8 keV.

Once the \lya\ (or \ion{He}{2}) luminosity is monotonically 
linked with the AGN X-ray luminosity,
we estimate the number density of AGN-powered sources
from the hard X-ray luminosity function \citep[e.g.,][]{Barger,Cowie}.
To power a blob with $L_{\rm Ly\alpha}$ = $10^{43}$ ${\rm ergs\ s^{-1}}$,
an AGN must have $L_X \ga 10^{41.8}$ ${\rm ergs\ s^{-1}}$,
which would generate a \heii blob with 
$L_{\rm He \sss II} \ga 10^{41.7}$ ${\rm ergs\ s^{-1}}$.
Around the required X-ray luminosity, 
\citet{Cowie} derive the number density of X-ray selected AGNs
regardless of their optical AGN signatures to be
$1.3 \times 10^{-5}$   $\lesssim$
$\Phi (L_X > 10^{42\ } {\rm ergs\ s^{-1}})$   $\ll$ 
$1.4\times10^{-4\ } {\rm Mpc^{-3}}$ at $2<z<4$. 
The extreme upper limit was determined by assigning all the unidentified
sources in the survey to $2<z<4$ and is thus very conservative.
For the brightest cooling sources in our simulations (Fig. \ref{fig:lf}),
we find 
$\Phi (L_{\rm Ly\alpha} \ga 10^{43})$ 
$\sim 3\times10^{-5}$ ${\rm Mpc^{-3}}$ and 
$\sim 9\times10^{-4}$ ${\rm Mpc^{-3}}$ 
for the condensed phase cut and 
the self-shielding correction cases, respectively.
\footnote{
The number density of \lya\ blobs (of unknown origin(s)) from the Subaru survey 
\citep{Matsuda} lies between these two cases.
}
Note again that our assumption that all the ionizing photons from all
AGNs are absorbed to produce \lya/\heii photons is very conservative
and that we are clearly over-predicting the number of AGN-powered blobs.
However, even under this conservative assumption
the number density of AGN-powered sources is only marginally
comparable to the number density of cooling sources in our simulations.
Therefore, at present, we conclude simply that a survey for 
extended \lya\ and \heii cooling radiation is not likely 
to be swamped by AGN-powered sources.

The above arguments are statistical, whereas 
distinguishing gravitational cooling radiation from the emission 
of AGN-photoionized gas for an \emph{individual} source 
requires a multi-wavelength approach.
It is therefore useful to target fields with a large amount of 
ancillary data (e.g., deep broad-band or X-ray imaging) 
to make an unambiguous detection of a true \heii cooling blob.
First, searching for the \ion{C}{4} (1549 \AA) emission line 
in the optical spectrum of the source is a good way 
to identify an AGN \citep[e.g.,][]{Keel}.
The composite spectra of optically selected quasars show bright
\ion{C}{4} lines, but much fainter \heii lines \citep{Telfer}.
The recently discovered \lya\ blob associated with a luminous 
mid-infrared source \citep{Dey} shows unusually strong \heha\ lines 
\emph{and} \ion{C}{4} lines in a localized region near the center of nebula, 
suggesting that this \lya\ blob is powered, at least in part, by an obscured AGN.
On the other hand, the absence of a \ion{C}{4} line in a spectrum with 
strong \heii emission might indicate gravitational cooling gas like that 
in our simulations.
As we discuss in the next section, the kinematics of the \heii line
can further constrain the origin of the \heii emission.
Second, if the AGN is heavily obscured, 
deep X-ray imaging of $\sim$ Msec will provide 
the most direct probe, because X-rays from the AGN can penetrate 
the large column densities of gas and dust.
%

\subsection{Superwinds}
\label{sec:superwind}
Alternatively, \citet{Taniguchi} suggest that galactic superwinds driven 
by starbursts could power the extended \lya\ blobs.
In this scenario, 
the collective kinetic energy of multiple supernovae is deposited
into the surrounding gas, producing a super-bubble filled with hot and high-pressure gas. 
If the mechanical energy overcomes the gravitational potential of the galaxies,
this metal-enriched gas blows out into the surrounding primordial IGM 
and evolves into superwinds.

While the luminosity and sizes of the observed blobs 
are roughly consistent with the predictions of simple wind models,
the mechanism to convert the mechanical
energy into \lya\ emission is not clear.
%
Using a fast-shock model, \citet{Francis} show 
that if the shocks are radiative,
the emission from the excited gas in the shock itself and 
the photoionized precursor region in front of the shock
can explain the observed \lya\ surface brightness of the blobs. 
For example, if we adopt the fiducial model of the pre-run shock grids
from \citet{Allen} \citep[MAPPINGS code by][]{Dopita} 
with a shock velocity of $700\ \mathrm{km\ s^{-1}}$, 
a number density of $10^{-2\ } \mathrm{cm^{-3}}$, 
a magnetic parameter of $B/\sqrt{n} = 2 \mu \mathrm{G\ cm^{3/2}}$, 
and solar metallicity,
then the \lya\ and \heha\ emissivity from the shock + precursor region will be 
$\sim 0.04$ and $0.002$ $\mathrm{ergs\ s^{-1}\ cm^{-2}}$, respectively.
If the shock is perpendicular to our line of sight, we would expect
surface brightnesses of $F_{\mathrm{Ly\alpha}} \sim 3 \times 10^{-18}$ and
$F_{\mathrm{He\sss II}} \sim 1.5 \times 10^{-19}\ \SB$ at $z=3$.
This \lya\ surface brightness is roughly consistent 
with the observed mean surface brightness of \lya\ blobs \citep{Matsuda},
but is an underestimate if we consider that \lya\ emission can be suppressed 
by various factors such as self-absorption and that the density of the IGM
is possibly lower than the assumed density.
If the IGM in the preshock region is composed of neutral primordial gas,
the UV photons produced in the post-shock plasma will ionize the preshock
region, and the lack of an effective cooling mechanism other than 
the atomic hydrogen 
and helium lines can boost the \lya\ and \heha\ line emissivities significantly.
However, the low metallicity shock grid is not currently available, and
the density of the IGM in the preshock region and the effect of mixing between
the metal-enriched winds and the pristine IGM are quite uncertain.
Thus it is difficult to predict how much mechanical energy is released 
through the \lya\ or \heha\ lines in the superwind model. 

One important feature of the superwind shock model is that
it also predicts many UV diagnostic lines (e.g., \ion{C}{4} $\lambda1549$)
that have been used to study the energetics of the narrow-line region in AGNs.
The debate about the origin of the \lya\ blobs
arises mainly because \lya\ is not a good diagnostic line
to discriminate between AGN photoionization and superwind shock-excitation
owing to its sensitivity to resonant scattering and obscuration by dust. 
Ideally, line ratios (e.g., \ion{He}{2}/\ion{C}{4}, once detected) 
could be used to discriminate among the different mechanisms.

The kinematics of the blob is potentially another test of the superwind hypothesis,
because of the expected bipolar outflow motion of the expanding shell.
For example, \citet{Ohyama} claim that Blob 1 of \citet{Steidel} shows
both blueshifted and redshifted components ($\sim \pm 3000\ \mathrm{km\ s^{-1}}$)
in the central region, and they attribute these profiles to the expanding bipolar
motion of a shocked shell driven by a superwind.
On the other hand, using integral field spectrograph data, 
\citet{Bower} argue that Blob 1 has chaotic velocity structures 
that can be explained by the interaction of slowly rising buoyant material 
with cooling gas in the cluster potential, and that 
a powerful collimated outflow alone 
appears inconsistent with the lack of velocity shear across the blob.

We show the \heha\ luminosity-weighted velocity dispersions of the gas particles
associated with individual blobs in Figure \ref{fig:veldist}.
The effect of Hubble expansion and peculiar motion 
is included in the velocity dispersion calculations 
but the thermal broadening for each gas particle is not.
Most halos have velocity dispersions smaller than $\sim$ 400 km s$^{-1}$, 
compared to the typical superwind speed of several hundreds to a 1000 km s$^{-1}$ 
\citep[e.g.,][]{Heckman,Pettini}.
For the superwind case, because the observed \lya\ emission comes mainly from 
the shock between the outflow from a galaxy and the surrounding pristine IGM, 
we expect the \heii emission to be as extended as the observed \lya\ emission.
Therefore, if we observe a spatially resolved \lya\ \emph{and} \heii emitting blob, 
and its velocity dispersion is larger than 400 km s$^{-1}$, 
it is possible to exclude cooling radiation as the source of that blob. 
Note that there would be no ambiguities in measuring the size and line 
broadening because \heha\ is optically thin.
Thus \heha\ is a finer tool than \hlya\ 
to study the kinematic properties of \lya\ blobs.

\section{CONCLUSIONS}
\label{sec:conclusion}
In this paper, we use high resolution cosmological simulations
to study the gravitational cooling lines arising from
gas accreted by forming galaxies.
Because baryons 
must radiate thermal energy to join a galaxy
and form stars, accreting gas produces extended 
\hlya\ emission (a ``\lya\ blob'') 
surrounding the galaxy.
We also expect cooling lines from singly ionized helium such as \heha\
to be present within \lya\ blobs.
We investigate whether three major atomic cooling lines, 
\hlya, \heha, and \helya\ are observable in the FUV and optical.
We discuss the best observational strategies to search for
cooling sources and how to distinguish them from other possible
mechanisms for producing \lya\ blobs.
Our principal findings are:

1. \hlya\ and \heha\ (\heii Balmer $\alpha$) cooling emission at $z=2-3$ 
are potentially detectable with deep narrow band imaging and/or 
spectroscopy from the ground.
\helya\ will be unreachable until a large aperture UV space 
telescope \citep[e.g. SUVO;][]{Shull} is available.

2. While our predictions for the strength of the \hlya\ emission line 
depend strongly on how to handle the self-shielded gas,
our predictions for the \heha\ line are rather robust 
owing to the negligible emissivity of \heii for the self-shielded IGM
below $T\sim10^{4.5\ }\K$.

3. Although \heha\ cooling emission is fainter than \lya\ 
by at least a factor of 10 and, unlike \lya\ blobs, 
might not be resolved spatially with current observational facilities, 
it is more suitable to study gas accretion in the galaxy formation process
because it is optically thin, less sensitive to the UV background, and
less contaminated by recombination lines from star-forming galaxies.

4. To use the \hlya\ and \heha\ cooling lines to constrain galaxy formation models, 
we first need to exclude the other possible mechanisms for producing \lya\ blobs.
First, because \heha\ emission from stars is limited to stars with very low metallicities 
($\mathrm{log}(Z/Z_{\sun}) \la -5.3$) and Population III objects,
its detection, unlike \lya, cannot be caused by stellar populations.
Second, the kinematics of the \heha\ line can distinguish gravitational cooling 
radiation from a scenario in which starburst-driven superwinds power \lya\ blobs,
because the \heii line width from cooling gas is narrower 
($\sigma < 400\ \mathrm{km s^{-1}}$) than the typical wind speeds 
(which are factors of several higher).
Third, if some fraction of the \heii emitting blobs are powered by AGN,
additional diagnostics such as the \ion{C}{4} line and/or X-ray emission
can be used to discriminate gravitationally cooling blobs from those powered, 
at least in part, by AGN. 

\acknowledgements
We thank an anonymous referee for his/her thorough reading of the 
manuscript and helpful comments. We thank Mark Fardal, Dusan Keres 
and Jane Rigby for valuable discussions.
YY and AIZ acknowledge funding from HST grant GO-09781.03-A, 
NASA LTSA grant NAG5-11108, and NSF grant AST 02-06084. 
NK and DHW were supported by NASA NAGS-13308 and NASA NNG04GK68G. 
In addition NK was supported by NSF AST-0205969.


\end{document}